\pdfoutput=1

\documentclass[11pt,twoside,a4paper,cmspaper,final,collab]{cms-tdr}
\def\svnVersion{175881}\def\cmsCernNo{2012-277}\def\cmsCernDate{\today}\def\cmsMessage{Submitted to the Journal of High Energy Physics}

\begin{document}\cmsNoteHeader{EXO-11-094}

\hyphenation{had-ron-i-za-tion}
\hyphenation{cal-or-i-me-ter}
\hyphenation{de-vices}

\RCS$Revision: 162495 $
\RCS$HeadURL: svn+ssh://svn.cern.ch/reps/tdr2/papers/EXO-11-094/trunk/EXO-11-094.tex $
\RCS$Id: EXO-11-094.tex 162495 2012-12-21 21:04:41Z ferencek $
\newlength\cmsFigWidth
\ifthenelse{\boolean{cms@external}}{\setlength\cmsFigWidth{0.85\columnwidth}}{\setlength\cmsFigWidth{0.4\textwidth}}
\ifthenelse{\boolean{cms@external}}{\providecommand{\cmsLeft}{top}}{\providecommand{\cmsLeft}{left}}
\ifthenelse{\boolean{cms@external}}{\providecommand{\cmsRight}{bottom}}{\providecommand{\cmsRight}{right}}
\cmsNoteHeader{EXO-11-094} % This is over-written in the CMS environment: useful as preprint no. for export versions
\title{Search for narrow resonances and quantum black holes\\in inclusive and b-tagged dijet mass
spectra\\from pp collisions at $\sqrt{s}=7$~TeV}

\newcommand{\cPWpr}{\ensuremath{\cmsSymbolFace{W}^\prime}\xspace}

\hyphenation{assuming}

\date{\today}

\abstract{
A search for narrow resonances and quantum black holes is performed in inclusive and b-tagged dijet mass spectra measured
with the CMS detector at the LHC. The data set corresponds to 5~fb$^{-1}$ of integrated luminosity collected in pp
collisions at $\sqrt{s}=7$~TeV. No narrow resonances or quantum black holes are observed. Model-independent upper limits
at the 95\% confidence level are obtained on the product of the cross section, branching fraction into dijets, and
acceptance for three scenarios: decay into quark-quark, quark-gluon, and gluon-gluon pairs. Specific lower limits are set
on the mass of string resonances (4.31~TeV), excited quarks (3.32~TeV), axi\-gluons and colorons (3.36~TeV),
scalar color-octet resonances (2.07~TeV), E$_{6}$ diquarks (3.75~TeV), and on the masses of W$^\prime$ (1.92~TeV) and
Z$^\prime$ (1.47~TeV) bosons. The limits on the minimum mass of quantum black holes range from 4 to 5.3~TeV. In addition,
b-quark tagging is applied to the two leading jets and upper limits are set on the production of narrow dijet resonances
in a model-independent fashion as a function of the branching fraction to b-jet pairs.
}

\hypersetup{%
pdfauthor={CMS Collaboration},%
pdftitle={Search for narrow resonances and quantum black holes in inclusive and b-tagged dijet mass
spectra from pp collisions at sqrt(s)=7 TeV},%
pdfsubject={CMS},%
pdfkeywords={CMS, physics, software, computing}}

\maketitle %maketitle comes after all the front information has been supplied

\section{Introduction}

Events with two or more energetic jets in the final state are copiously produced in proton-proton (pp) collisions at the
Large Hadron Collider (LHC). Such events arise when the constituent partons are scattered with large transverse momenta
\pt. The invariant mass spectrum of the dijet system, consisting of the two jets with the largest \pt (leading jets), is
predicted by quantum chromodynamics (QCD) to fall steeply and smoothly. However, there are numerous extensions of the
standard model (SM) that predict the existence of new massive particles that couple to quarks (\cPq) and gluons (\cPg),
and result in the appearance of resonant structures in the dijet mass spectrum. Furthermore, the dijet mass spectrum
can be used to search for quantum black holes. Hence, dijet events provide one of the event topologies used to
search for new physics.

In this Letter we report on a search for narrow resonances in the inclusive dijet mass spectrum measured with the Compact
Muon Solenoid (CMS) detector at the LHC in pp collisions at a center-of-mass energy of 7~TeV. We complement the generic
search with a more flavour-specific analysis, in which information based on displaced secondary vertices is used to
identify jets resulting from the hadronization and decay of a \cPqb\ quark. As a consequence, the analysis has an enhanced
sensitivity to objects that decay preferentially into \bbbar pairs.

Although the results of this search are applicable to any massive narrow resonance decaying to two jets, we consider
specific models predicting the following narrow $s$-channel dijet resonances:
\begin{itemize}
\item String resonances (S), which are Regge excitations of quarks and gluons in string theory and decay predominantly
to \cPq\cPg~\cite{Anchordoqui:2008di,Cullen:2000ef}.
\item Scalar diquarks (D), which decay to $\cPq\cPq$ and \cPaq\hspace{1 pt}\cPaq, predicted by a grand unified theory based on the
E$_6$ gauge symmetry group~\cite{ref_diquark}.
\item Mass-degenerate excited quarks (\cPq$^*$), which decay to \cPq\cPg, predicted in quark compositeness models~\cite{ref_qstar,Baur:1989kv};
the compositeness scale is set to be equal to the mass of the excited quark.
\item Axial-vector particles called axigluons (A), which decay to \cPq\cPaq, predicted in a model where the symmetry
group SU(3) of QCD is replaced by the chiral symmetry $\text{SU(3)}_\text{L} \times
\text{SU(3)}_\text{R}$~\cite{ref_axi}.
\item Color-octet colorons (C), which also decay to \cPq\cPaq; these are vector particles predicted by the flavour-universal
coloron model, in which the SU(3) gauge symmetry of QCD is embedded in a larger gauge group~\cite{ref_coloron}.
\item Scalar color-octet resonances (S8)~\cite{Han:2010rf} that appear in many dynamical electroweak symmetry breaking
models such as Technicolor. We consider the decay channel into a pair of gluons.
\item Massive scalar color-octet resonances (S8$_\text{\cPqb}$)~\cite{Bai:2010dj} that result from the breaking of an
$\text{SU(3)}\times\text{SU(3)}$ gauge symmetry down to the QCD gauge group and that may have generically large
couplings to $\cPqb$ quarks. We consider the production of a coloron that subsequently decays into an
S8$_\text{\cPqb}$ and a light scalar singlet. We fix the singlet mass to 150~GeV.
The S8$_\text{\cPqb}$ and scalar singlet have branching fractions ($B$) of approximately
$100\%$ to \bbbar and \cPg\cPg, respectively. The tangent of the mixing angle $\theta$ between the two
SU(3) gauges is set to 0.15. This resonance search is inclusive of extra jets, so the search strategy is insensitive to
the decay of the low-mass singlet state.
\item New gauge bosons (\cPWpr and \cPZpr), that decay to \cPq\cPaq, predicted by models that include new gauge
symmetries~\cite{ref_gauge}; the \cPWpr and \cPZpr\ bosons are assumed to have standard-model-like couplings.
Consequently, the ratio between the branching fraction of the \cPZpr\ to \bbbar and the branching fraction to a
pair of quarks (excluding the top quark) is approximately 0.22.
\item Randall--Sundrum (RS) gravitons (G), which decay to \cPq\cPaq\ and \cPg\cPg, predicted in the RS model of extra
dimensions~\cite{ref_rsg}. The value of the dimensionless coupling $k/\overline{M}_\text{Pl}$ is chosen to be
0.1, where $k$ is the curvature scale in the 5-dimensional anti de Sitter space and $\overline{M}_\text{Pl}$ is the
reduced Planck scale. The ratio between the branching fraction of the RS graviton to \bbbar and the branching fraction
to a pair of quarks (excluding the top quark) or gluons is approximately 0.1~\cite{Allanach:2002gn}.
\end{itemize}

In addition, we report on a search for quantum black holes~\cite{MR,Calmet,qbh1} in the inclusive dijet mass spectrum.
This search is motivated by theories with low-scale quantum gravity, which offer a novel solution to the hierarchy problem
of the standard model by lowering the scale of quantum gravity $M_\text{D}$ from the Planck scale ($M_\text{Pl}\sim 10^{16}$~TeV) to
a lower value $M_\text{D}\sim1$~TeV, i.e. a value of the order of the electroweak symmetry breaking scale.
Examples of models using this approach are the Arkani-Hamed--Dimopoulos--Dvali (ADD) model~\cite{add,add1} and the
Randall--Sundrum (RS) model~\cite{RS,ref_rsg}. In the former model, extra dimensions are flat and compactified
on a torus or a sphere, while in the latter model, a single extra dimension ($n = 1$) is warped. The strengthened gravity
allows for formation of quantum black holes with masses $M_\text{QBH}$ close to the quantum gravity scale $M_\text{D}$.
Such objects evaporate faster than they thermalize, resulting in a non-thermal decay into a pair of jets, rather than a
high-multiplicity final state~\cite{Calmet,qbh1}. An earlier search for quantum black holes performed by the CMS experiment~\cite{QBH2011}
was based on an analysis of high-multiplicity, energetic final states.

The searches presented in this document exceed the sensitivity to new physics of previous
CMS~\cite{Khachatryan:2010jd,Chatrchyan:2011ns,QBH2011} and ATLAS~\cite{ATLAS2010,Aad:2011aj,Aad:2011fq} published
searches. A summary of recent searches for dijet resonances and a comparison of the approaches between different
experiments are presented in Ref.~\cite{Harris:2011bh}. The most recent dedicated search for \bbbar resonances in the dijet
final state at a hadron collider was performed by the CDF experiment in Run~I of the Tevatron~\cite{Abe:1998uz}.

\section{The CMS detector and data sample}

The CMS experiment uses a right-handed coordinate system, with the origin at the center of the detector. The $z$-axis points along the
direction of the counterclockwise beam and the x-axis points to the centre of the LHC; $\phi$ is the
azimuthal angle, covering $-\pi<\phi\leq\pi$, $\theta$ is the polar angle, and the pseudorapidity $\eta \equiv -\ln[\tan(\theta/2)]$.

The central feature of the CMS apparatus is a superconducting solenoid of 6~m internal diameter providing an
axial magnetic field of 3.8~T. Within the field volume in the central pseudorapidity region are the silicon-pixel and
silicon-strip tracker ($|\eta| < 2.4$) and the barrel and endcap calorimeters ($|\eta| < 3$) consisting of a lead-tungstate
crystal electromagnetic calorimeter (ECAL) and a brass/scintillator hadron calorimeter (HCAL). An iron/quartz-fibre
Cherenkov calorimeter is located in the forward region ($3 < |\eta| < 5$), outside the field volume. For triggering purposes and
to facilitate jet reconstruction, the ECAL and HCAL cells are grouped into towers projecting radially outward from the
center of the detector. The energy deposits measured in the ECAL and the HCAL within each projective tower are summed to
obtain the calorimeter tower energy. A more detailed description of the CMS detector, including its muon subdetectors,
can be found elsewhere~\cite{Adolphi:2008zzk}.

The integrated luminosity of the data sample used for this analysis is
$4.98\pm0.11$~fb$^{-1}$~\cite{CMS-PAS-SMP-12-008}, and corresponds to the full data sample recorded by the CMS experiment
in 2011. Events are recorded using a two-tier trigger system. The sample was collected using a combination of triggers requiring
the presence of jets in the event. At the start of the data-taking period, a multijet trigger based on \HT was used, where
\HT is the scalar sum of the transverse momenta of all jets in the event with \pt above 40 GeV. Over the course of the
data-taking period, the \HT threshold of the lowest unprescaled \HT trigger was increased from 350 to 750~GeV to keep the
overall trigger rate approximately constant as the number of additional pp collisions in the same or adjacent bunch crossings
(pileup interactions) was increasing. To mitigate the negative impact of increasing \HT thresholds on the overall trigger
efficiency, a dedicated dijet-mass trigger based on ``wide-jet'' reconstruction, the offline reconstruction technique described
in Section~\ref{sc:evt_reco_sel}, was introduced toward the end of the data-taking period. Events with dijet masses greater
than 850~GeV and pseudorapidity separation between the two jets $|\Delta\eta|<2$ are selected online with this dedicated trigger.
The efficiency of all of the triggers used in this analysis is measured from the data to be larger than 99.8\% for dijet
masses above 890~GeV.

\section{Event reconstruction and selection
\label{sc:evt_reco_sel}}

Events selected by the trigger system are required to be consistent with coming from a pp collision and have at least one
reconstructed primary vertex within $\pm$24~cm of the detector center along the beam line and within 2~cm of the
detector center in the plane transverse to the beam.

Jets are reconstructed offline using the anti-\kt clustering algorithm~\cite{Cacciari:2008gp} with a distance
parameter of 0.5. The four-momenta of particles reconstructed by the CMS particle-flow (PF)
algorithm~\cite{CMS-PAS-PFT-09-001,CMS-PAS-PFT-10-001} are used as input to the jet-clustering algorithm.
The particle-flow algorithm combines information from all CMS subdetectors to provide a complete list of
long-lived particles in the event. Reconstructed and identified particles include muons, electrons (with associated
bremsstrahlung photons), photons (including conversions in the tracker volume), and charged and neutral hadrons. The
reconstructed jet energy $E$ is defined as the scalar sum of the energies of the constituents of the jet, and the jet
momentum $\vec{p}$ as the vector sum of their momenta. The jet transverse momentum \pt is the component
of $\vec{p}$ perpendicular to the beam. All reconstructed jets used in this analysis are required to pass identification
criteria that are fully efficient for signal events~\cite{CMS-PAS-JME-10-003}, in order to remove possible instrumental
and non-collision backgrounds in the selected sample. The mising transverse energy \MET is defined as the magnitude of
the vector sum of the transverse momenta of all particles reconstructed in the event.

The jet energy scale is calibrated using jet energy corrections derived from Monte Carlo simulation, test beam
results, and collision data~\cite{Chatrchyan:2011ds}. The corrections account for extra energy clustered into jets from
pileup interactions on an event-by-event basis~\cite{Cacciari:2007fd}.
Additional corrections for the flavor of the jet are small (${<}1\%$) and are not applied; however,
when b tagging is applied, the systematic uncertainty in the jet energy scale is increased to account for the
different fragmentation and decay properties of heavy-flavor-originated jets.

Calibrated PF jets are clustered into what are called ``wide jets''~\cite{Chatrchyan:2011ns}. The wide jet
reconstruction technique, inspired by performance studies of different jet definitions~\cite{Cacciari:2008gd}, increases
the search sensitivity by recombining large-angle final-state QCD radiation from the outgoing partons, resulting in an
improved dijet mass resolution. The clustering starts with the two leading jets, which are both required to have
$|\eta|<2.5$. No explicit requirement on \pt of the two leading jets is applied. All other
jets with $\pt>30$~GeV and $|\eta|<2.5$ are added to the closest leading jet if they are within $\Delta
R\equiv\sqrt{(\Delta\eta)^2+(\Delta\phi)^2}<1.1$, where $\Delta\eta$ and $\Delta\phi$ are the distances between the two
jets in $\eta$ and $\phi$, respectively. In this way two wide jets are formed. Compared to
our previous search~\cite{Chatrchyan:2011ns}, the minimum \pt threshold for subleading jets used in the wide-jet technique
has been increased from 10~GeV to 30~GeV in order to be more robust against jets coming from pileup interactions.

The dijet system is composed of the two wide jets. We require that the pseudorapidity separation $\Delta\eta$ of the
two wide jets satisfies $|\Delta\eta|<1.3$, and that both wide jets be in the region $|\eta|<2.5$. These requirements
maximize the search sensitivity for isotropic decays of dijet resonances in the presence of QCD background.
The dijet mass is given by $m=\sqrt{(E_1 + E_2)^2 - (\vec{p}_1 + \vec{p}_2)^2}$, where $E_1$ ($E_2$) and $\vec{p}_1$
($\vec{p}_2$) are the energy and momentum of the leading (next-to-leading) jet. For the trigger selection to be
fully efficient, we select events with $m>890$~GeV without any requirement on wide-jet \pt. To study possible impact of pileup
on the analysis, the rate of selected events, defined as the number of events passing the event selection per unit of
integrated luminosity, over the course of the data-taking period was analyzed. Despite the increasing pileup, the rate
of selected events was found to be stable.

Jets from the hadronization and decay of $\cPqb$ quarks are identified ("tagged") by the characteristically long lifetime of B
hadrons. The combined-secondary-vertex (CSV) algorithm~\cite{CMS-PAS-BTV-11-004} uses variables from reconstructed
secondary vertices together with track-based lifetime information to distinguish jets that originate from a $\cPqb$
quark from those that originate from lighter quarks and gluons. This algorithm was tuned for $\cPqb$ jets from top-quark
decays but shows good performance in other types of events as well. Based on a study of the expected upper limits and the properties
of the tagger, the loose operating point of the CSV tagger was chosen for this analysis. The ratio of the tagging
efficiency between data and simulation is measured in a b-quark-enriched sample~\cite{CMS-PAS-BTV-11-004}. This
data-to-simulation ``scale factor'' is found to depend on the jet \pt, but it is close to unity (within ${\sim}5$\%). A
similar scale factor is measured for light jets passing the b-tagging criteria ("mistags") and is found to depend on the
jet \pt and $\eta$, but it is also close to unity (within ${\sim}10$\%). Because of the limited number of jets at high \pt,
the scale factors are measured up to a jet \pt of 670~GeV and are extrapolated to higher values of the jet \pt. To take into
account additional uncertainty associated with the extrapolation procedure, larger uncertainties are assigned to the
extrapolated values of the scale factors. Only the leading subjet in each of the two wide jets is considered for $\cPqb$
tagging. Therefore, events can be separated into three exclusive categories: 0, 1, and 2 b tags.

\section{Measurement of the dijet mass spectrum
\label{sc:dijet_mass}}

The dijet mass spectrum used to search for narrow dijet resonances is defined as
\begin{linenomath}
\begin{equation}
 \frac{\text{d}\sigma}{\text{d}m}\simeq\frac{1}{\int L \text{d}t}\frac{N_i}{\Delta m_i},
 \label{eq:dijet_spectrum}
\end{equation}
\end{linenomath}
where $m$ is the dijet mass, $N_i$ is the number of events in the $i$-th dijet mass bin, $\Delta m_i$ is the width of
the $i$-th dijet mass bin, and $\int L \text{d}t$ is the integrated luminosity of the data sample. The size of dijet mass
bins is approximately equal to the dijet mass resolution~\cite{Khachatryan:2010jd}. To test the smoothness of the
measured dijet mass spectrum, we fit the following parameterization to the data:
\begin{linenomath}
\begin{equation}
 \frac{\text{d}\sigma}{\text{d}m}=\frac{P_0(1-m/\sqrt{s})^{P_1}}{(m/\sqrt{s})^{P_2+P_3\ln(m/\sqrt{s})}},
 \label{eq:bkg_parameterization}
\end{equation}
\end{linenomath}
where $P_0$, $P_1$, $P_2$, and $P_3$ are free parameters and $\sqrt s = 7$~TeV. This functional form has been used in previous
searches~\cite{Aaltonen:2008dn,ATLAS2010,Khachatryan:2010jd,Aad:2011aj} to describe both data and QCD predictions.

\begin{figure}[!htb]
 \centering
 \begin{tabular}{cc}
   \includegraphics[width=0.48\textwidth]{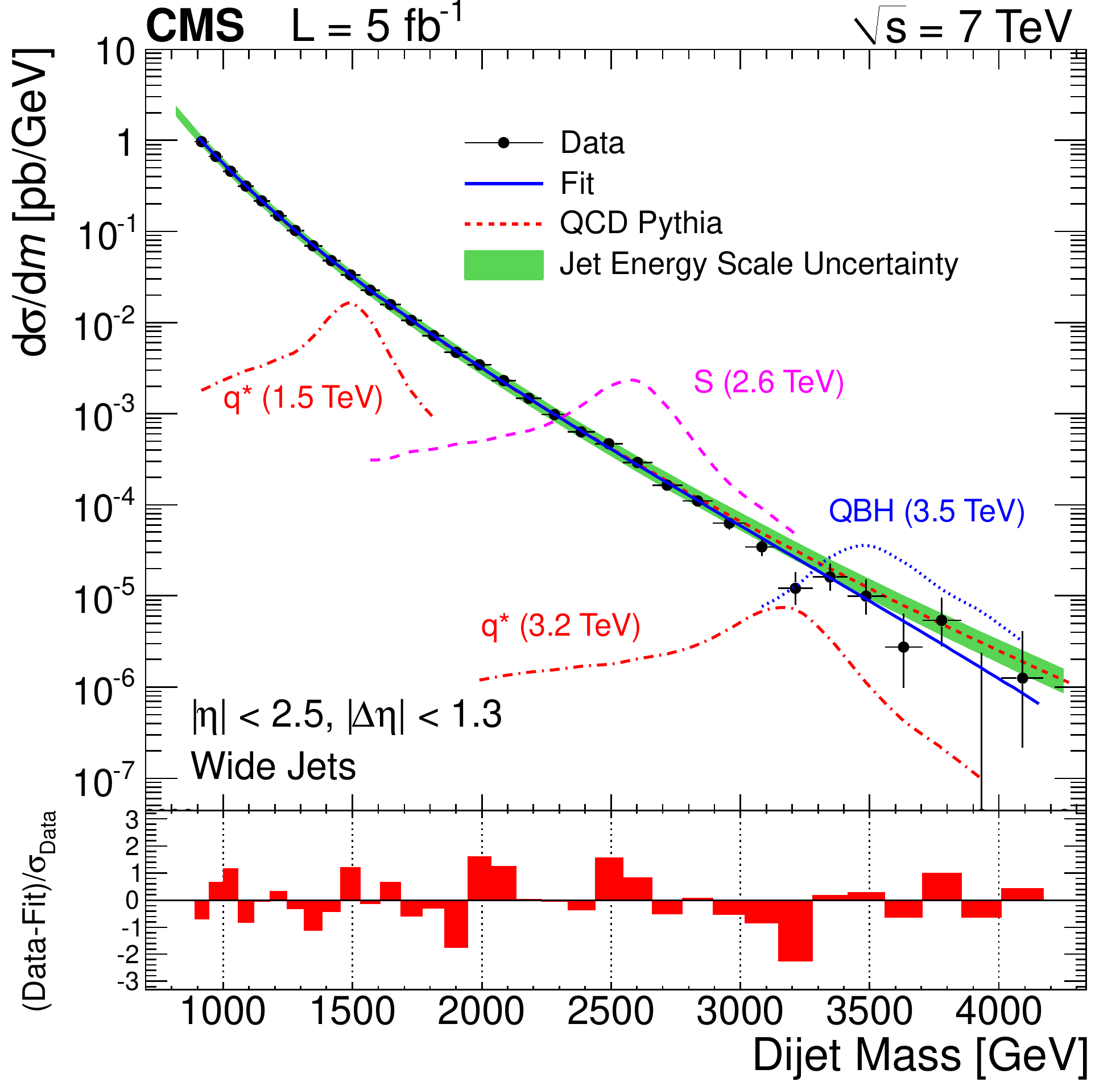} &
   \includegraphics[width=0.48\textwidth]{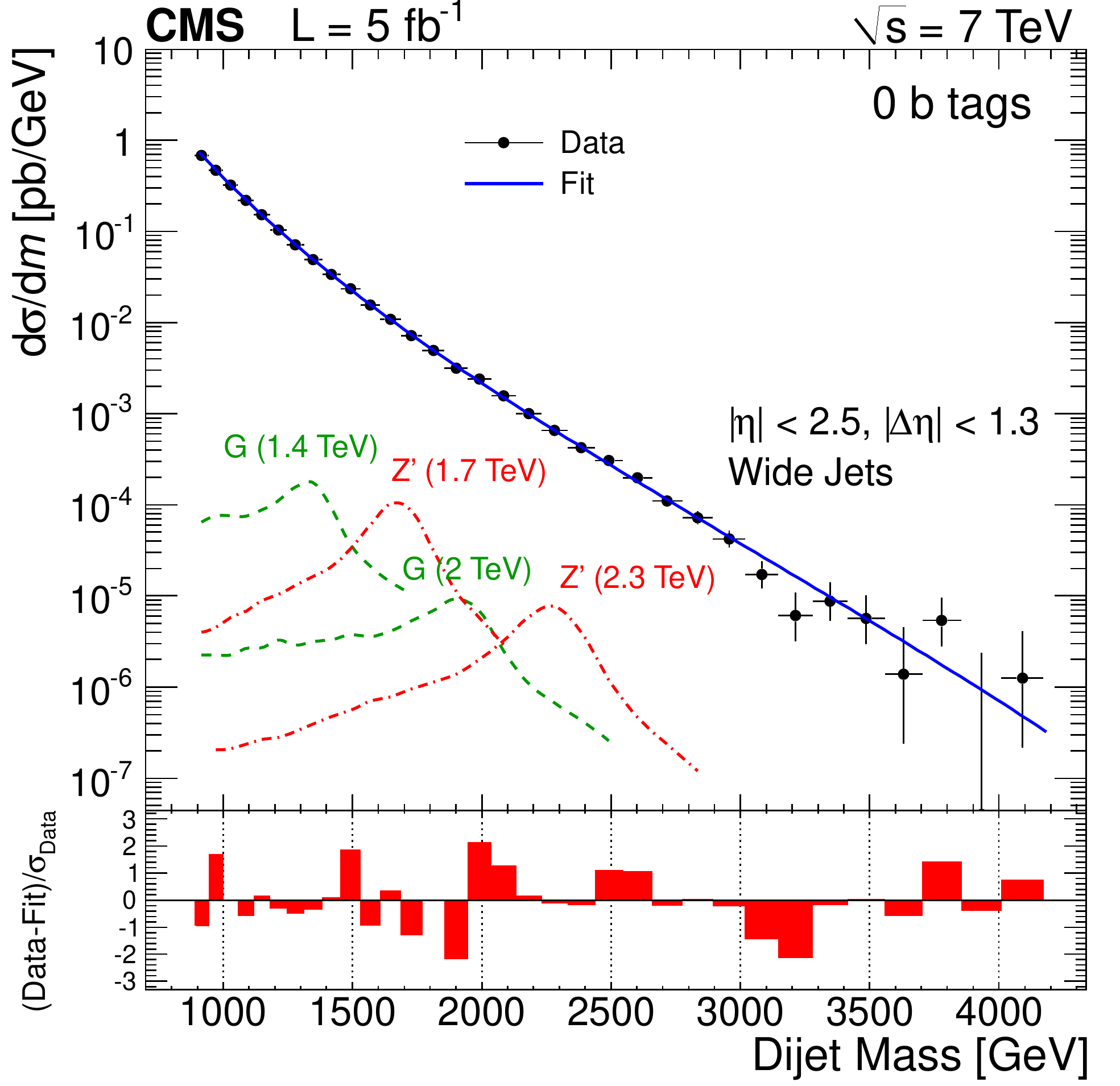} \\
   (a) & (b) \\
   \includegraphics[width=0.48\textwidth]{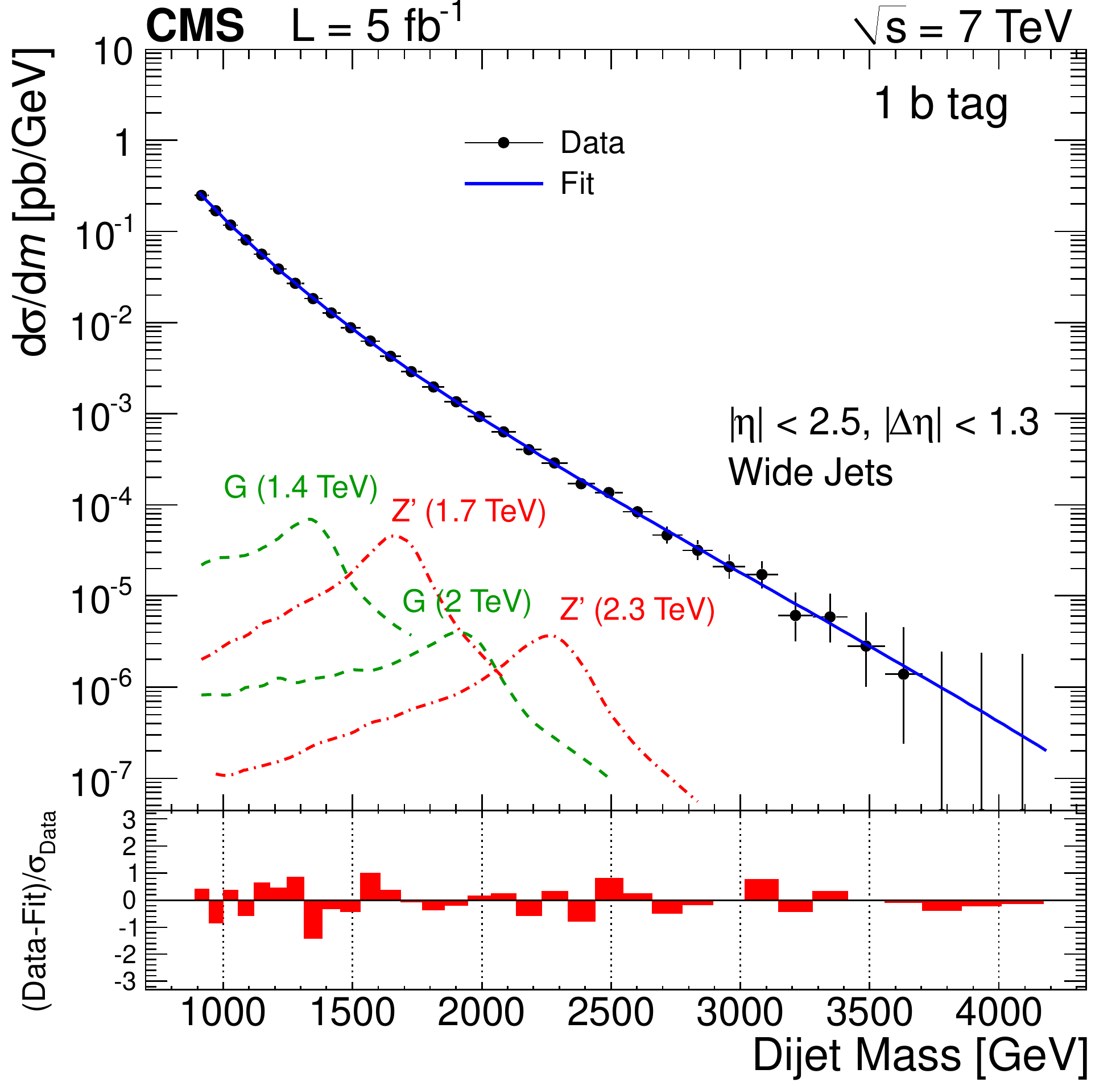} &
   \includegraphics[width=0.48\textwidth]{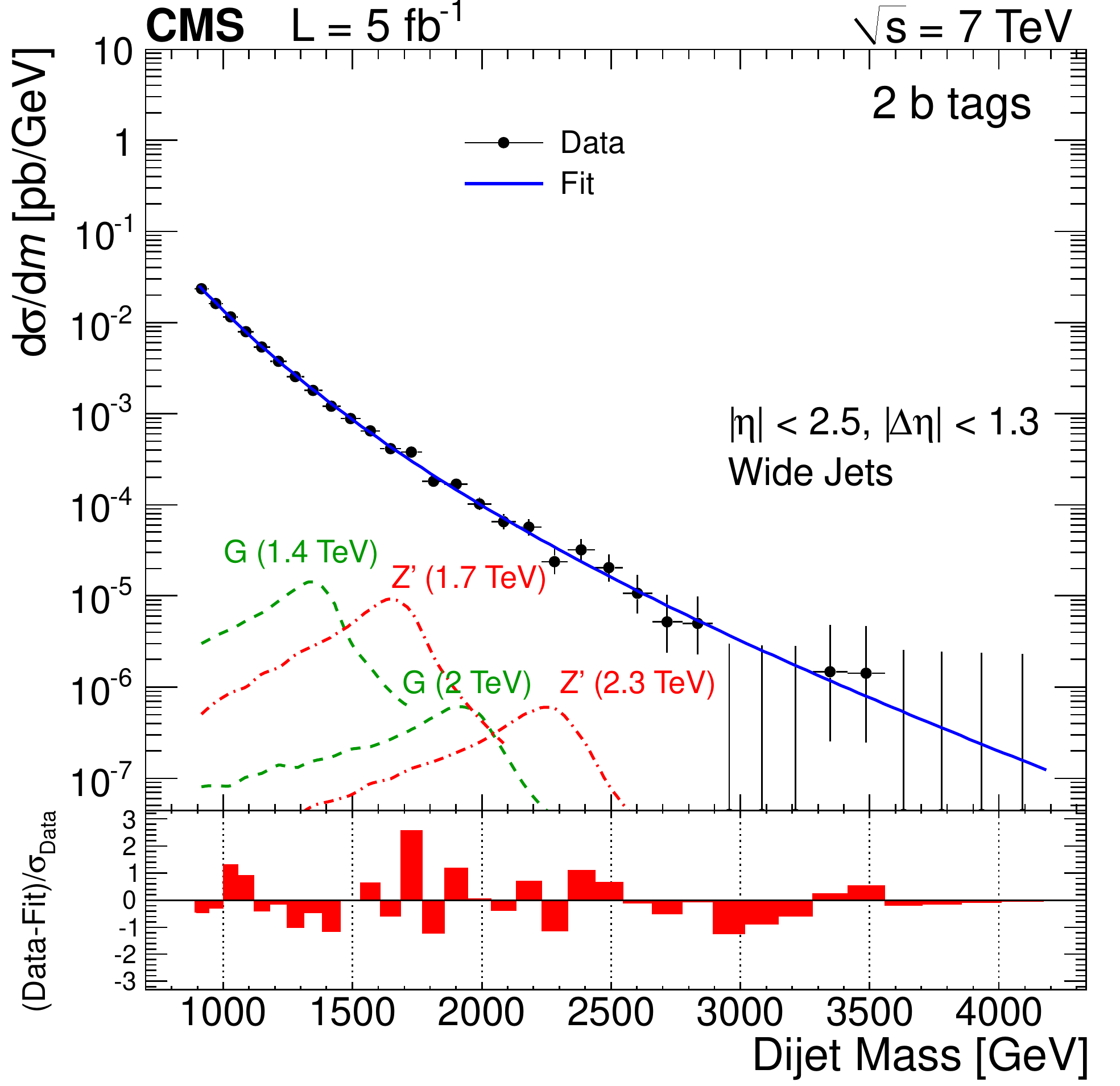} \\
   (c) & (d)
 \end{tabular}
 \caption{(a) Inclusive dijet mass spectrum from wide jets (points) compared to a smooth fit (solid) and
    predictions for QCD (short-dashed), excited quarks (\cPq$^*$), string resonances (S),
    and quantum black holes (QBH). The QCD prediction has been normalized to the data (see text).
    The shaded band shows the contribution from the systematic uncertainty in the jet energy scale.
    (b), (c), (d) Dijet mass spectra from wide jets (points) for different b-tag multiplicity categories compared to a
    smooth fit (solid) and predictions for RS graviton (G) and \cPZpr. The bin-by-bin fit residuals
    are shown at the bottom of each plot. The functional form of the fit is described in the text.
 \label{fig:dijet_mass_spectra}}
\end{figure}

Figure~\ref{fig:dijet_mass_spectra} (a) presents an inclusive dijet mass spectrum for the two wide jets, a fit to the
data, and bin-by-bin fit residuals, defined as the difference between the data and the fit value divided by the
statistical uncertainty in the data. The vertical error bars are central intervals with correct coverage for Poisson
variation, and the horizontal error bars are the bin widths. The data are compared to a QCD prediction from
\PYTHIA~\cite{Sjostrand:2006za} ({\sc v6.4.24}), which includes a simulation of the CMS detector based on
\GEANTfour~\cite{Agostinelli:2002hh,Allison:2006ve} ({\sc v4.9.4}) and the jet energy corrections. The prediction uses a
renormalization scale $\mu=\pt$ of the hard-scattered partons with the CTEQ6L1 parton distribution functions (PDFs)~\cite{Pumplin:2002vw}
and the Z2 underlying event tune (the Z2 tune is identical to the Z1 tune~\cite{Field:2010bc} except
that Z2 uses the CTEQ6L1 PDFs), and has been normalized to the data by multiplying the prediction by a factor of 1.22.
This factor was derived by scaling the number of predicted events with $m>890$~GeV to that observed in data.
The shape of the leading-order (LO) QCD prediction is in agreement with the data.
Figures~\ref{fig:dijet_mass_spectra} (b), (c), and (d) present the dijet mass spectra, fits to the data, and the
bin-by-bin fit residuals for the three b-tag multiplicity categories: 0, 1, and 2 b tags.

Based on the fit residuals and the values of the reduced $\chi^2$ obtained, no significant deviations from
the fit function are observed in the measured dijet mass spectra, indicating that the data are well described by a
smooth function.

\section{Search for narrow dijet resonances and quantum black holes
\label{sc:search}}

We search for narrow dijet resonances, for which the natural resonance width is small compared to the CMS dijet mass
resolution~\cite{Harris:2011bh}, and for quantum black holes. The dijet mass shape of narrow dijet resonances depends
primarily on the type of partons coming from the resonance decay, because this affects both the amount of radiation and
the response of the detector to final state jets. Using \PYTHIA and the CMS detector simulation, the dijet mass
shapes for the following parton pairings are predicted: $\cPq\cPaq$ (or $\cPq\cPq$) resonances from the process
$\cPG\rightarrow \cPq\cPaq$~\cite{ref_rsg}, $\bbbar$ resonances from $\cPG\rightarrow\bbbar$~\cite{ref_rsg},
$\cPq\cPg$ resonances from $\cPq^* \rightarrow \cPq\cPg$~\cite{ref_qstar}, and $\cPg\cPg$ resonances from $\cPG
\rightarrow \cPg\cPg$~\cite{ref_rsg}. The predicted dijet mass shapes have a Gaussian core coming from
the jet energy resolution, and a tail towards lower mass arising from QCD radiation and steeply falling parton
distribution functions. The dijet mass shapes are relatively narrow for $\cPq\cPaq$ ($\cPq\cPq$) resonances, wider for
$\bbbar$ and $\cPq\cPg$ resonances, and are the widest for $\cPg\cPg$ resonances. The increase of the width of the
measured mass shape and the shift of the mass distribution towards lower masses are enhanced when the number of gluons
in the final state is larger, because gluons are more likely to radiate than quarks. The dijet mass shapes are wider for
$\bbbar$ resonances because of the presence of neutrinos from the semileptonic b decays that escape detection.

It is commonly assumed~\cite{Dimopoulos:2001hw,Giddings:2001bu} that the minimum mass of quantum black holes
$M_\text{QBH}^\text{min}$ cannot be smaller than $M_\text{D}$. However, the formation
threshold can be significantly larger than $M_\text{D}$. For a given $M_\text{D}$, the dijet mass shapes for quantum
black holes are fairly independent of the number of extra dimensions $n$ and would appear as bumps in a steeply
falling QCD dijet mass spectrum, as shown in Fig.~\ref{fig:dijet_mass_spectra} (a). The dijet mass shapes for quantum
black holes are modeled using the {\sc qbh} ({\sc v1.03}) matrix-element generator~\cite{qbh_gen} with the CTEQ6L PDF
set~\cite{Pumplin:2002vw}, followed by the parton showering simulation with \PYTHIA and a fast parametric simulation of the CMS
detector~\cite{FastSim}.

\begin{figure}[!htb]
 \centering
 \begin{tabular}{cc}
   \includegraphics[width=0.48\textwidth]{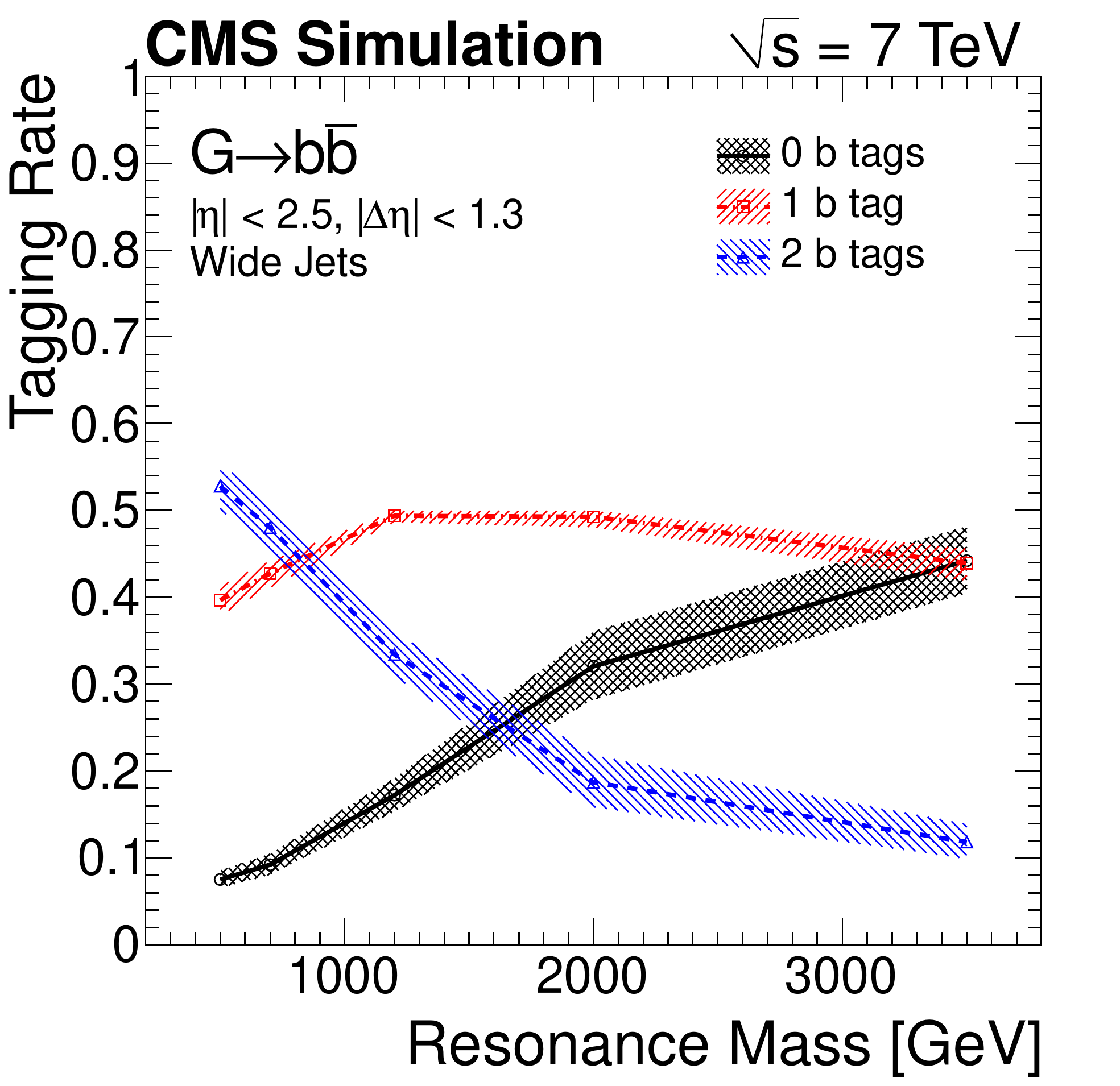} &
   \includegraphics[width=0.48\textwidth]{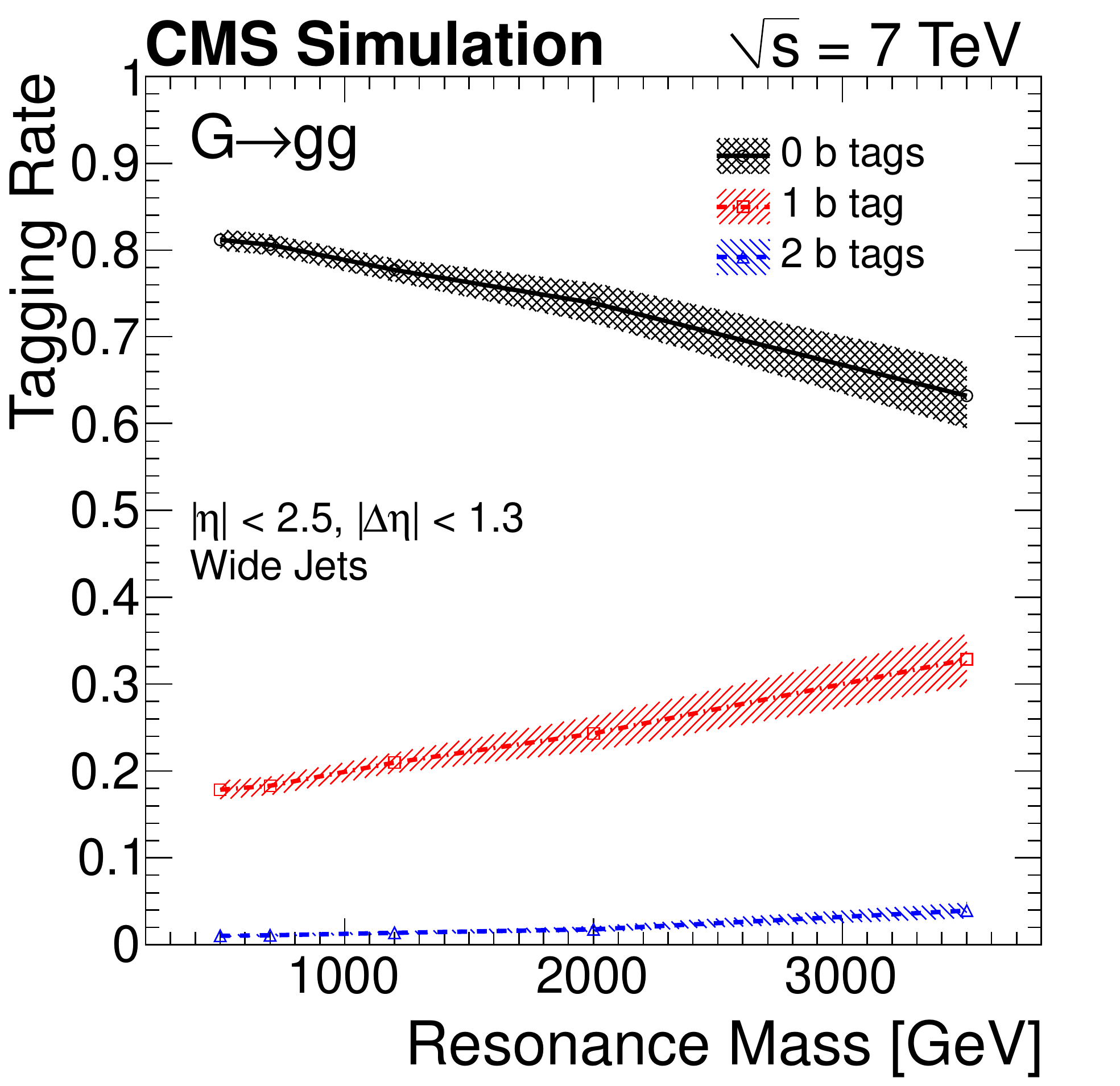} \\
   (a) & (b)
 \end{tabular}
 \caption{Tagging rates for $0$, $1$, and $2$ b tags as a function of the resonance mass for (a) \bbbar and (b)
   $\cPg\cPg$ decay modes of the RS graviton (G). Hatched regions represent uncertainties in the rates due to
   variations of the b-tag scale factors within their uncertainties. The tagging rates for the $\cPq\cPaq$
   ($\cPq=\cPqu,\cPqd,\cPqs$) decay modes are similar to the $\cPg\cPg$ tagging rates.
 \label{fig:eff_curves}}
\end{figure}

Based on the number of $\cPqb$-tagged jets, events are separated into three exclusive categories: 0-, 1-, and 2-tag categories.
The tagging rate for each of these categories is defined as the fraction of events ending up in that category.
The tagging rates as a function of the resonance mass are derived for different decay modes of RS gravitons and are shown
in Fig.~\ref{fig:eff_curves} for the \bbbar and $\cPg\cPg$ decay modes. As can be seen in the figure, the efficiency to
correctly tag a $\cPqb$ jet decreases as the resonance mass increases. The rate of double-tagging a resonance that
decays into two light quarks or gluons remains below ${\sim}5\%$ throughout the mass range. The tagging rates
for the $\cPq\cPaq$ ($\cPq=\cPqu,\cPqd,\cPqs$) decay modes are similar to the $\cPg\cPg$ tagging rates. The rate of
double-tagging a resonance that decays into two charm quarks is systematically higher than for light flavor decay modes but
is still significantly lower than for the \bbbar decay mode (by a factor of ${\sim}4$ at a resonance mass of 1~TeV).
Rather than introduce an additional dependence of the result on the branching fraction to $\cPqc$ quarks, we assume that the
$\cPqc\cPaqc$ decay mode has the same tagging rates as the light quark and gluon decay modes. This assumption simplifies the
interpretation of the analysis by removing an extra parameter at the cost of slightly reduced sensitivity.
The tagging rates shown in Fig.~\ref{fig:eff_curves} are assumed to be universally applicable to all narrow
resonances decaying into the same type of partons. 

Since the tagging rates for all non-\bbbar decay modes are (conservatively) assumed to be the same, the only free
parameter that specifies the fraction of 0, 1, and 2 b-tag events originating from a narrow resonance with a given mass
is the \bbbar decay fraction $f_{\bbbar}$ defined at the parton level as
\begin{linenomath}
\begin{equation}
 f_{\bbbar} = \frac{B(\text{X}\rightarrow\bbbar)}{B(\text{X}\rightarrow\text{jj})},
 \label{eq:bbbar_fraction}
\end{equation}
\end{linenomath}
where X is a generic narrow resonance. As $f_{\bbbar}$ increases, the fraction of events from a resonance populating
the 2 b-tag spectrum is expected to increase, depending on the tagging rates shown in Fig.~\ref{fig:eff_curves}.
Because of the inefficiency in b tagging, even in the case of $f_{\bbbar}=1$, a fraction of events still populates the 0 and 1 b-tag
spectra.

\subsection{Statistical treatment and systematic uncertainties
\label{sc:stat}}

No significant deviations from the expected background have been observed in the measured dijet mass spectra. We use the measured dijet
mass spectra, the background parameterization, and the dijet mass shapes to set upper limits on
$\sigma\times B\times A$, the product of the production cross section ($\sigma$), branching fraction
($B$) for the jet-jet final state, and acceptance ($A$) for the kinematic requirements $|\eta|<2.5$ and
$|\Delta\eta|<1.3$. The acceptance for isotropic decays is $A\approx 0.6$, independent of the heavy resonance mass.

For setting upper limits, we use a Bayesian formalism~\cite{Beringer:1900zz} with a flat prior on the signal cross section,
consistent with other dijet resonance searches at the LHC~\cite{Chatrchyan:2011ns,Aad:2011fq}; log-normal priors are
used to model systematic uncertainties, which are marginalized as nuisance parameters. We calculate the posterior
probability density as a function of resonance cross section independently at each value of the resonance mass. With b
tagging applied, the data from each of the three tagged spectra are combined into a single likelihood to provide a
single limit by assuming a particular value for $f_{\bbbar}$.

In order to achieve good coverage properties for the confidence intervals in the presence of a signal that is not yet strong
enough to be observed, the data are fit to the background function plus a signal line shape with the signal cross
section treated as a free parameter. The resulting fit function with the signal cross section set to zero is used as the
background hypothesis. The uncertainty in the background fit is incorporated by marginalizing over the background fit
parameters (not including the signal cross section) after diagonalizing the covariance matrix to account for the
correlations in the parameters. We also calculate the expected upper limits on $\sigma\times B\times A$ using
pseudo-experiments: ensembles of simulated experiments generated from the smooth background parameterization obtained
from the signal-plus-background fit to the data.

While events from a resonance that are double-tagged are dominated by the \bbbar final state ({assuming} that $f_{\bbbar}$
is not trivially small), there remains an ambiguity for the 0 and 1 b-tag cases. Resonances such as the RS graviton
decay into pairs of gluons as well as $\cPq\cPaq$ pairs. On the other hand, particles such as the \cPZpr\ or S8$_\text{\cPqb}$
decay exclusively into $\cPq\cPaq$ final states. Because of the gluon's larger color factor, gluons radiate more
than quarks, resulting in a broader dijet mass shape and, consequently, weaker expected limits. While the wide-jet
reconstruction technique mitigates this effect, the limits depend on whether the 0 and 1 b-tag mass
shapes are dominated by gluons or quarks in the final state. Therefore, when b tagging is applied,
two sets of upper limits are placed on $\sigma\times B\times A$, one for resonances that decay into
gluons in addition to b quarks (``gg/bb'') and one for resonances that decay into quarks only (``qq/bb'').
Mass shapes appropriate to gg or qq resonances are used in conjuction with a bb mass shape used for both
types of resonances. The mass shapes in each tag category are weighted according to the expected gluon, quark, or
$\cPqb$-quark content, as determined by the tagging rates and $f_{\bbbar}$.

In the inclusive analysis, the dominant sources of systematic uncertainty are the jet energy scale (2.2\%), the jet
energy resolution (10\%), the integrated luminosity determination (2.2\%)~\cite{CMS-PAS-SMP-12-008}, and the statistical uncertainty
in the background parameterization, where the uncertainties in the sources are given in parentheses. The statistical uncertainty
in the background parameterization leads to the uncertainty in the expected background yields, with the double-tagged dijet
mass spectrum having the largest uncertainty that ranges from ${\sim}1\%$ at a dijet mass of 1~TeV to ${\sim}15\%$ at 3.5 TeV. The jet
energy scale and the resolution uncertainties are incorporated into the limit-setting calculation by marginalizing over
nuisance parameters that control the mean and the width of the dijet mass shape. For the b-tagged analysis, the
uncertainties in the b-tag scale factors (${\sim}5\%$ for heavy and ${\sim}10\%$ for light flavor
jets)~\cite{CMS-PAS-BTV-11-004} are also considered. The flavor dependence of the energy response for PF jets at high jet
\pt (${>}100$~GeV) relevant for this analysis is well within the jet energy scale uncertainty~\cite{Chatrchyan:2011ds};
nevertheless, for the b-tagged analysis, the jet energy scale uncertainty is conservatively assigned to be 3\% for all
resonance masses considered.

\section{Results}

\begin{figure}[!htb]
  \centering
  \includegraphics[width=0.48\textwidth]{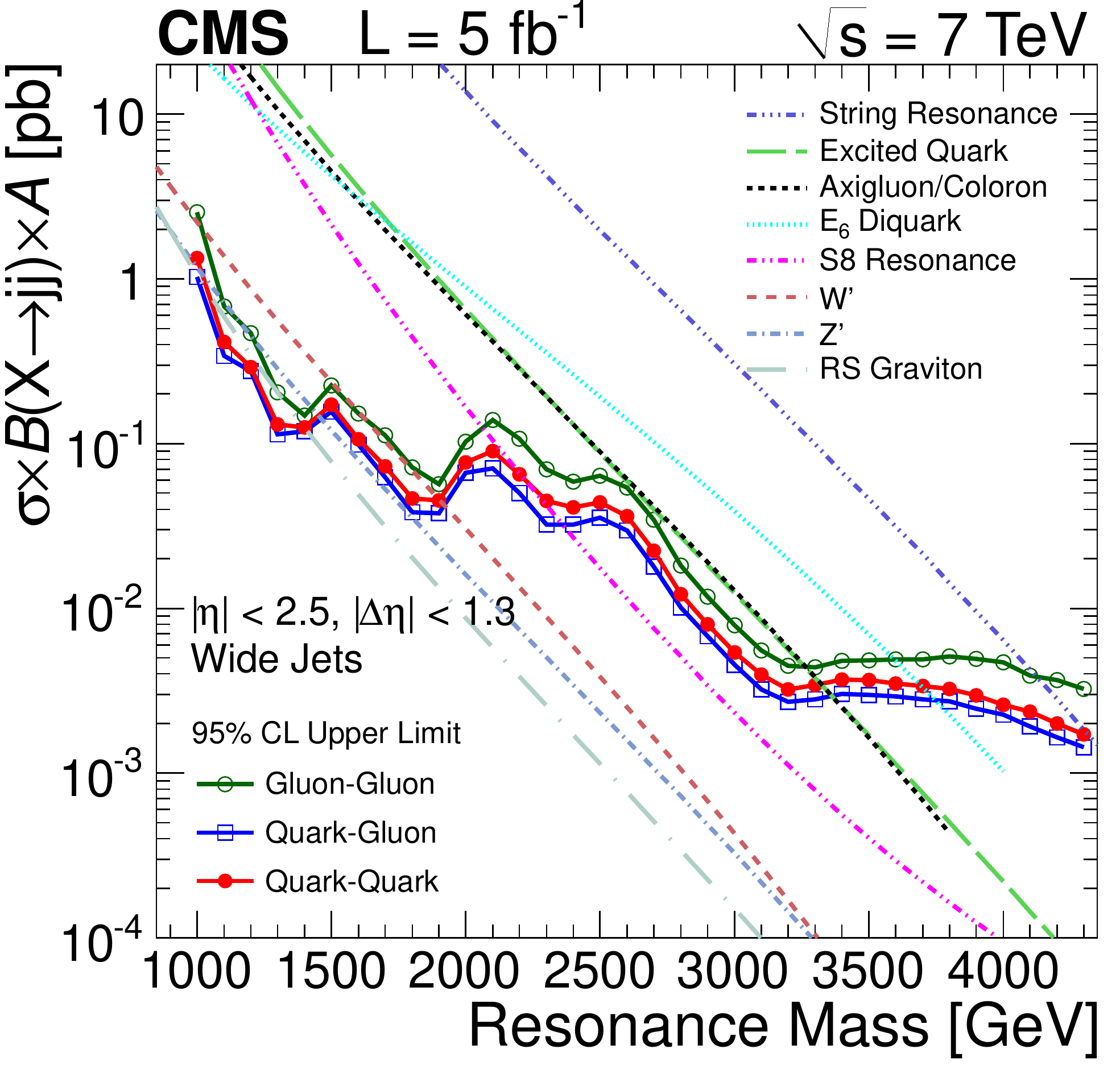}
  \caption{Observed 95\% CL upper limits on $\sigma\times B\times A$ for dijet resonances of type gluon-gluon
    (open circles), quark-gluon (solid circles), and quark-quark (open boxes) from the inclusive analysis, compared to
    theoretical predictions for string resonances~\cite{Anchordoqui:2008di,Cullen:2000ef}, excited
    quarks~\cite{ref_qstar,Baur:1989kv}, axigluons~\cite{ref_axi,Chivukula:2011ng}, colorons~\cite{ref_coloron},
    E$_6$ diquarks~\cite{ref_diquark}, S8 resonances~\cite{Han:2010rf}, $\cPWpr$ and $\cPZpr$
    bosons~\cite{ref_gauge}, and Randall--Sundrum gravitons~\cite{ref_rsg}.
  \label{fig:limits_obs_inclusive}}
\end{figure}

Figure~\ref{fig:limits_obs_inclusive} shows the observed upper limits at the 95\% confidence level (CL) on
$\sigma\times B\times A$ for qq, qg, and gg resonances from the inclusive analysis. The observed upper limits
for signal masses between 1.0 and 4.3~TeV are also reported in Table~\ref{tab:limits_obs_inclusive}. The observed upper
limits can be compared to predictions of $\sigma\times B\times A$ at the parton level, without any detector
simulation, in order to determine mass limits on new particles. The theoretical predictions are obtained at LO with narrow
width approximation using CTEQ6L1 parton distribution functions~\cite{Pumplin:2002vw}. For S8$_\text{\cPqb}$ resonances,
a LO cross section is obtained using the \MADGRAPH matrix-element generator~\cite{Alwall:2011uj}. For axigluons and colorons,
we also take into account the next-to-leading-order $K$-factors~\cite{Chivukula:2011ng}. New particles are excluded at the
95\% CL in mass regions for which the theory curve lies above the upper limit for the appropriate pair of partons.

\begin{table}[!htbp]
 \caption{Observed 95\% CL upper limits on $\sigma\times B\times A$ for narrow quark-quark (\cPq\cPq), quark-gluon (\cPq\cPg)
   and gluon-gluon (\cPg\cPg) resonances with masses between 1.0 and 4.3~TeV, derived from an inclusive analysis of
   dijet mass spectra.
 \label{tab:limits_obs_inclusive}}
 \centering
 \begin{tabular}{|c|c|c|c||c|c|c|c|} \hline
   Mass   &  \multicolumn{3}{c||}{Upper limit on $\sigma\times B\times A$ [pb]} & Mass & \multicolumn{3}{c|}{Upper limit on $\sigma\times B\times A$ [pb]} \\
   $\,$[TeV]$\,$  & ~~~~~qq~~~~~   & ~~~~~qg~~~~~   & ~~~~~gg~~~~~  & $\,$[TeV]$\,$ & ~~~~~qq~~~~~ & ~~~~~qg~~~~~ & ~~~~~gg~~~~~ \\ \hline
   1.0 & 1.0    & 1.3    & 2.5    & 2.7 & 0.018  & 0.022  & 0.035 \\
   1.1 & 0.34   & 0.42   & 0.68   & 2.8 & 0.010  & 0.012  & 0.018 \\
   1.2 & 0.28   & 0.29   & 0.47   & 2.9 & 0.0068 & 0.0080 & 0.0118 \\
   1.3 & 0.11   & 0.13   & 0.20   & 3.0 & 0.0045 & 0.0054 & 0.0079 \\
   1.4 & 0.12   & 0.13   & 0.15   & 3.1 & 0.0032 & 0.0039 & 0.0056 \\
   1.5 & 0.16   & 0.17   & 0.23   & 3.2 & 0.0027 & 0.0032 & 0.0045 \\
   1.6 & 0.10   & 0.11   & 0.15   & 3.3 & 0.0028 & 0.0034 & 0.0044 \\
   1.7 & 0.062  & 0.073  & 0.112  & 3.4 & 0.0030 & 0.0037 & 0.0048 \\
   1.8 & 0.038  & 0.046  & 0.072  & 3.5 & 0.0030 & 0.0037 & 0.0048 \\
   1.9 & 0.038  & 0.045  & 0.057  & 3.6 & 0.0029 & 0.0035 & 0.0049 \\
   2.0 & 0.066  & 0.077  & 0.103  & 3.7 & 0.0028 & 0.0034 & 0.0049 \\
   2.1 & 0.071  & 0.090  & 0.139  & 3.8 & 0.0027 & 0.0032 & 0.0051 \\
   2.2 & 0.050  & 0.065  & 0.107  & 3.9 & 0.0025 & 0.0030 & 0.0049 \\
   2.3 & 0.032  & 0.045  & 0.070  & 4.0 & 0.0023 & 0.0026 & 0.0047 \\
   2.4 & 0.032  & 0.041  & 0.059  & 4.1 & 0.0019 & 0.0024 & 0.0039 \\
   2.5 & 0.035  & 0.044  & 0.064  & 4.2 & 0.0016 & 0.0020 & 0.0037 \\
   2.6 & 0.030  & 0.036  & 0.054  & 4.3 & 0.0014 & 0.0017 & 0.0032 \\
 \hline
 \end{tabular}
\end{table}

\begin{figure}[!htb]
 \centering
 \begin{tabular}{cc}
   \includegraphics[width=0.48\textwidth]{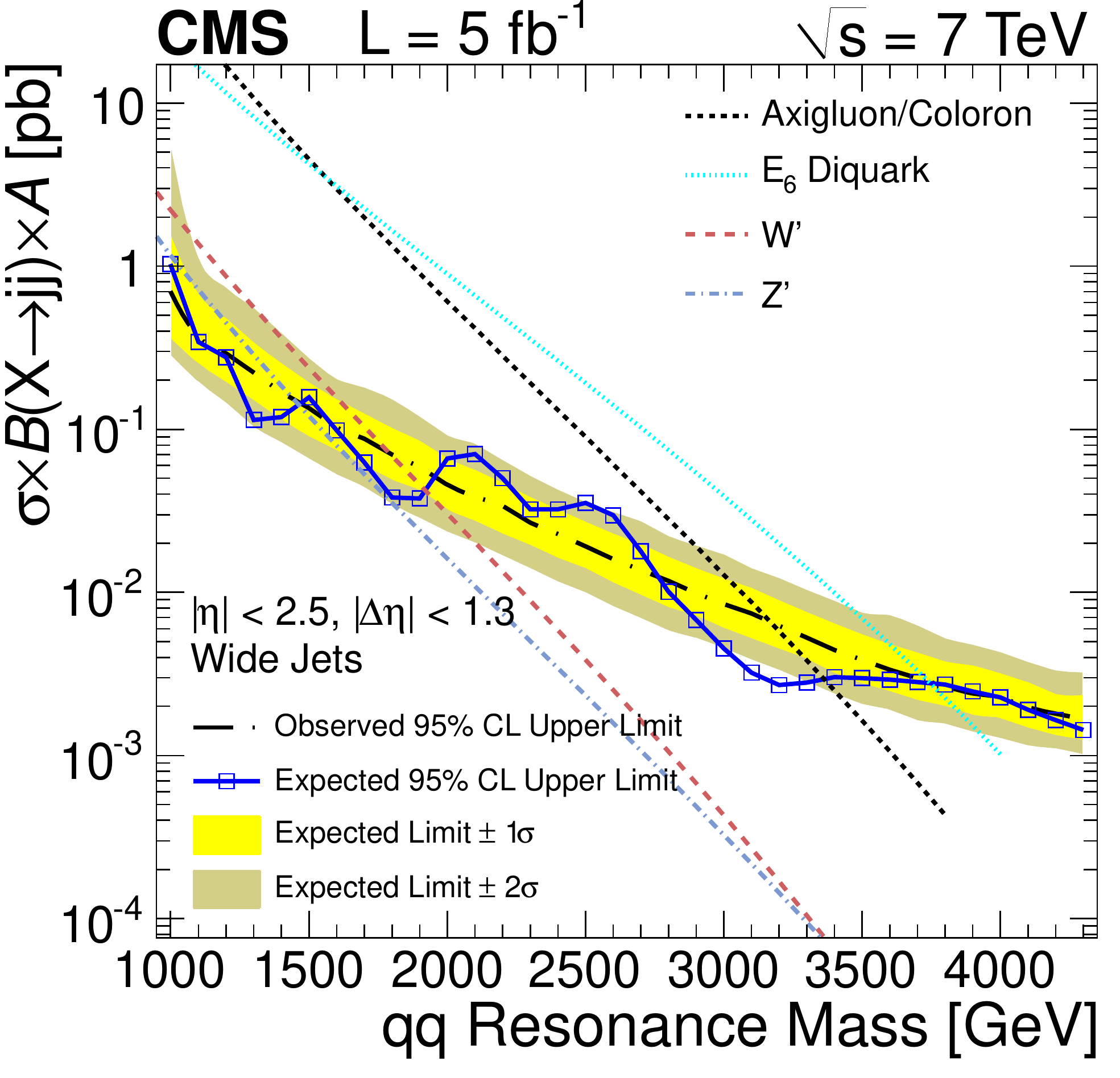} &
   \includegraphics[width=0.48\textwidth]{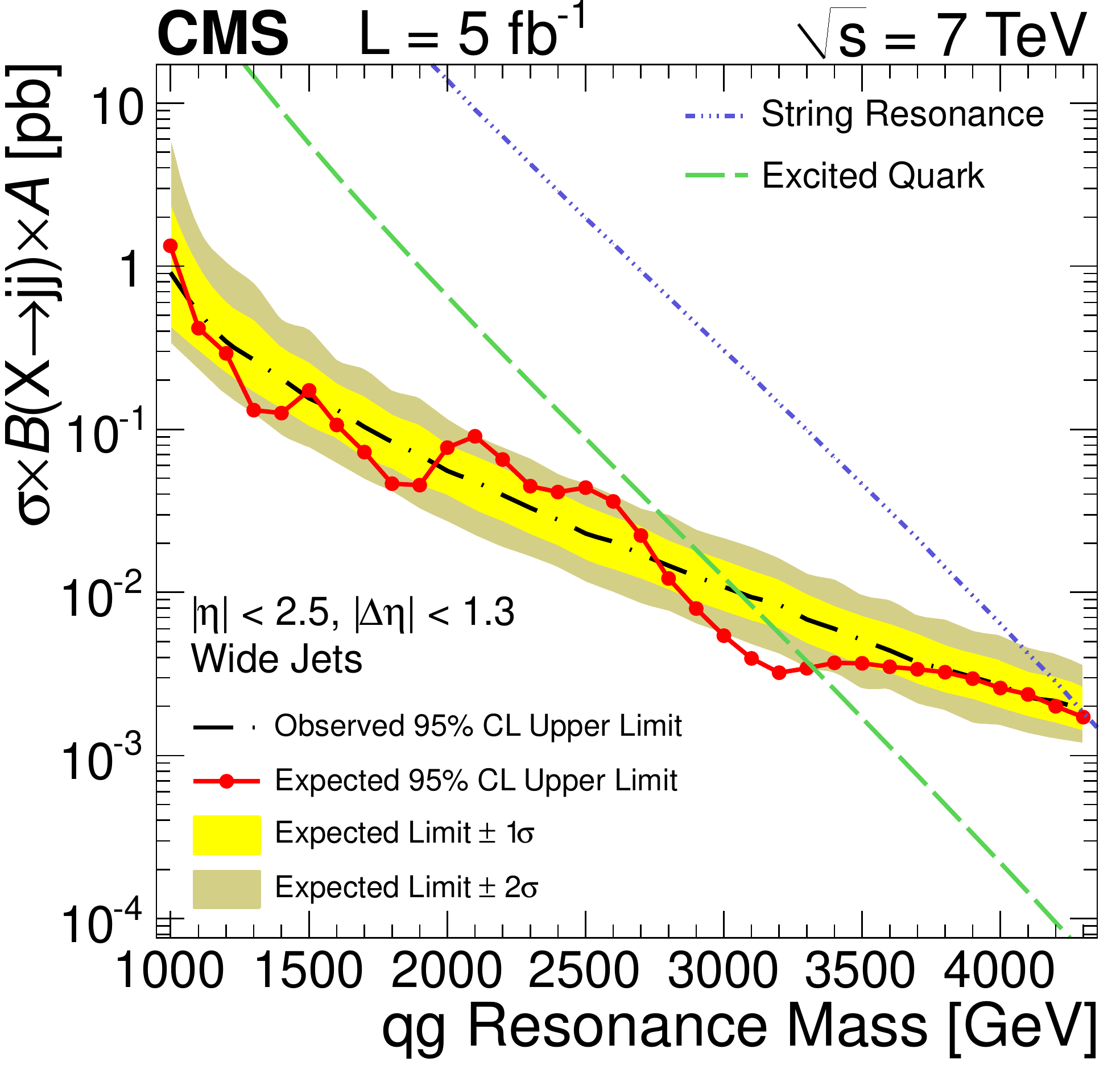} \\
   (a) & (b) \\
   \multicolumn{2}{c}{\includegraphics[width=0.48\textwidth]{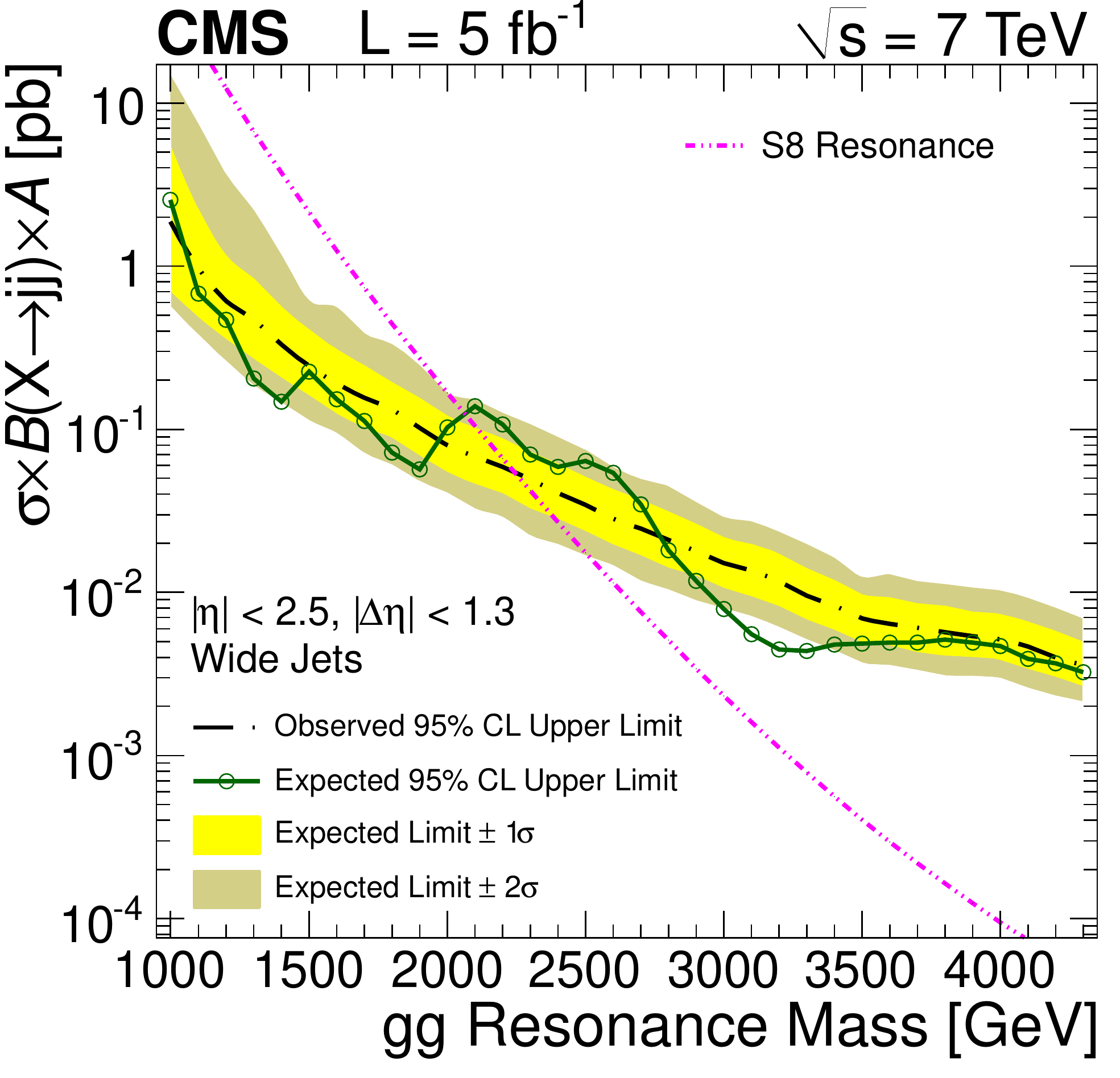}} \\
   \multicolumn{2}{c}{(c)}
 \end{tabular}
 \caption{Observed 95\% CL upper limits on $\sigma\times B\times A$ for (a) quark-quark, (b) quark-gluon, (c) and
   gluon-gluon dijet resonances (points) from the inclusive analysis are compared to the expected limits (dot-dashed) and
   their variation at 1$\sigma$ and 2$\sigma$ levels (shaded bands). Theoretical predictions for various resonance
   models are also shown.
 \label{fig:limits_exp_inclusive}}
\end{figure}

Figure~\ref{fig:limits_exp_inclusive} shows the expected limits and their uncertainty bands for qq, qg, and gg resonances
and compares them to both the observed limits and theoretical predictions for new resonances. Upward fluctuations in
data observed around 2 and 2.5 TeV result in observed limits that are less stringent than the expected ones;
conversely, a downward fluctuation around 3.2 TeV results in more stringent observed limits than the expected ones.

For string resonances, we exclude masses smaller than $4.31$~TeV; this extends our previous exclusion of
$0.5 < M(\text{S}) < 4.0$~TeV~\cite{Khachatryan:2010jd, Chatrchyan:2011ns}.
For excited quarks, we exclude masses smaller than $3.32$~TeV; this extends our previous exclusion of $0.5 < M(\cPq^*) <
2.49$~TeV~\cite{Khachatryan:2010jd, Chatrchyan:2011ns} and extends the ATLAS exclusion at $2.99\TeV$~\cite{Aad:2011fq}.
For E$_6$ diquarks, we exclude masses in the range $1.0 < M(\text{E}_6) < 3.75$~TeV; this extends our previous exclusion
at $3.52$~TeV~\cite{Khachatryan:2010jd, Chatrchyan:2011ns, Harris:2011bh}. For axi\-gluons or colorons, we exclude
masses smaller than $3.36$~TeV; this extends our previous exclusion of $0.50<M(\text{A},\text{C})<2.47$~TeV~\cite{Khachatryan:2010jd, Chatrchyan:2011ns}
and is similar to the ATLAS limit of $3.32\TeV$ based on 1~fb$^{-1}$ of data~\cite{Aad:2011fq}. (We note here that the ATLAS and
CMS experiments use different methods to calculate the axigluon and coloron cross section, which results in noticeable
differences in the expected and observed mass limits for these models~\cite{Harris:2011bh}.)

For the S8 color-octet model, we exclude masses in the range $1.0 < M(\mbox{S8}) < 2.07$~TeV; this extends the previous ATLAS exclusion of
$0.9 < M(\text{S8}) < 1.92\TeV$~\cite{Aad:2011fq}. For $\cPWpr$ bosons, we exclude masses in the range $1.00<M(\cPWpr)<1.92$~TeV;
this extends the previous CMS exclusion limit $1.0<M(\cPWpr)<1.51$~TeV~\cite{Khachatryan:2010jd, Chatrchyan:2011ns}.
Finally, we exclude $\cPZpr$ bosons in the mass range $1.0<M(\cPZpr)<1.47$~TeV. The observed and expected mass exclusions
for specific models of dijet resonances are summarized in Table~\ref{tab:mass_exclusions_inclusive} and are in generally good agreeement.

With the present data set, we start to be sensitive to the Randall--Sundrum gravitons just above 1~TeV of mass. For the
specific case of the Randall--Sundrum graviton, which couples either to a pair of gluons or to a quark-antiquark pair,
the model-dependent limits on cross section are derived using a weighted average of the $\cPq\cPaq$ and $\cPg\cPg$ dijet
mass shapes, where the weights correspond to the relative branching fractions for these two final states. Although not
strictly correct, approximate limits can be obtained by defining the model-dependent limits as a weighted average of the
model-independent $\cPq\cPq$ and $\cPg\cPg$ limits. In the case of the Randall--Sundrum graviton, this approximate procedure was found to
produce upper limits that differ by as much as 20\% from those obtained using the weighted dijet mass shapes. However, for
steeply falling signal cross sections, this difference would result in a relatively modest difference in the mass limit. 

\begin{table}[!htb]
  \caption{Observed and expected 95\% CL mass exclusions for specific models of dijet resonances from the inclusive
    analysis.}
  \centering
  \normalsize
  \begin{tabular}{|c|c|c|c|}
  \hline
    Model & Final State & Exp. Mass Exclusion & Obs. Mass Exclusion \\
          &             & [TeV]     & [TeV] \\
    \hline
    String Resonance (S)       & $\cPq\cPg$  & [1.0, 4.29] & [1.0, 4.31] \\
    Excited Quark (\cPq$^*$)   & $\cPq\cPg$  & [1.0, 3.05] & [1.0, 3.32] \\
    E$_6$ Diquark (D)          & $\cPq\cPq$  & [1.0, 3.74] & [1.0, 3.75] \\
    Axigluon (A) / Coloron (C) & $\cPq\cPaq$ & [1.0, 3.16] & [1.0, 3.36] \\
    S8 Resonance (S8)          & $\cPg\cPg$  & [1.0, 2.24] & [1.0, 2.07] \\
    $\cPWpr$ Boson ($\cPWpr$)  & $\cPq\cPaq$ & [1.0, 1.78] & [1.0, 1.92] \\
    $\cPZpr$ Boson ($\cPZpr$)  & $\cPq\cPaq$ & [1.0, 1.45] & [1.0, 1.47] \\
    \hline
  \end{tabular}
\label{tab:mass_exclusions_inclusive}
\end{table}

\begin{figure}[!htb]
 \centering
 \begin{tabular}{cc}
   \includegraphics[width=0.48\textwidth]{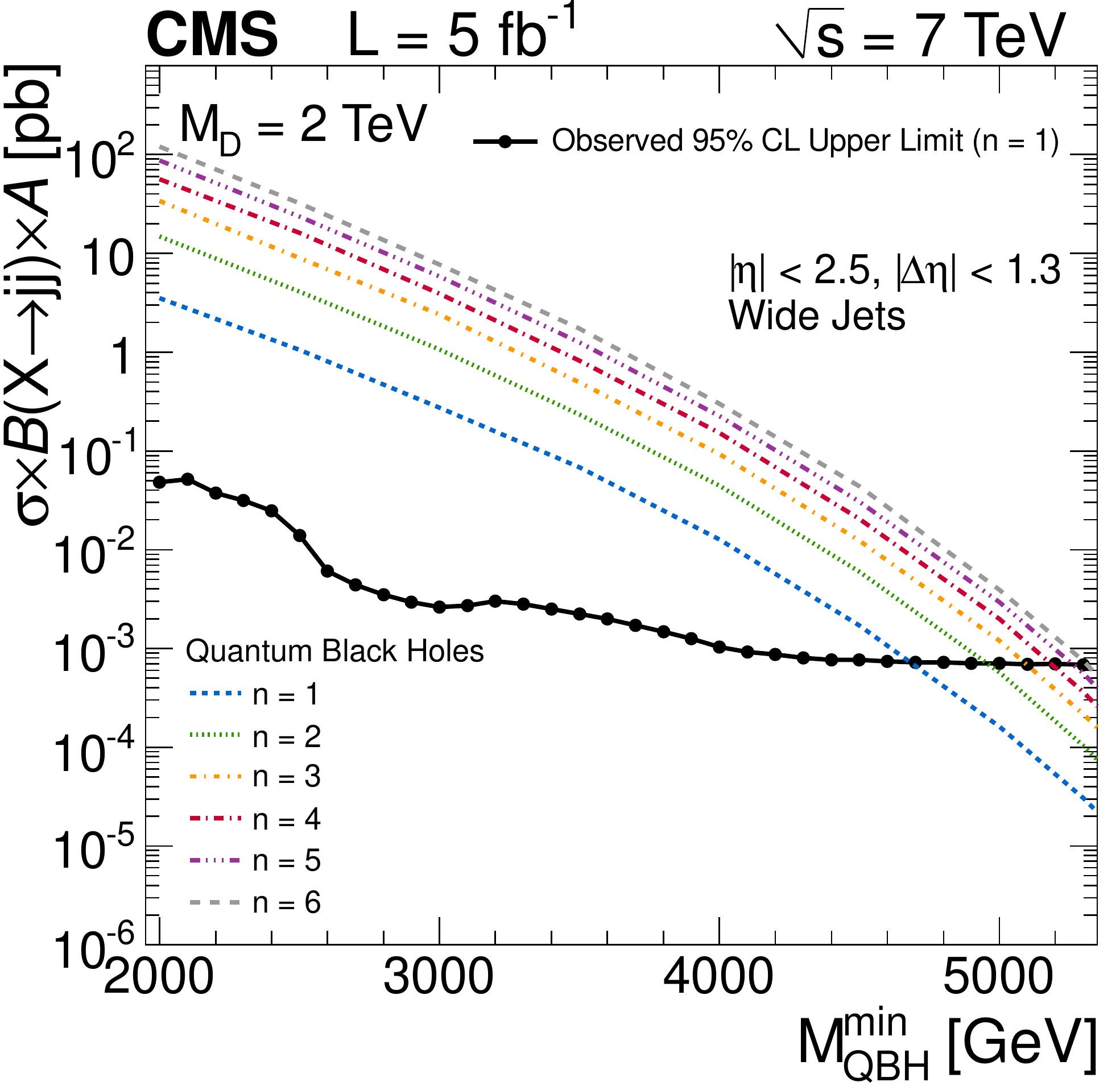} &
   \includegraphics[width=0.48\textwidth]{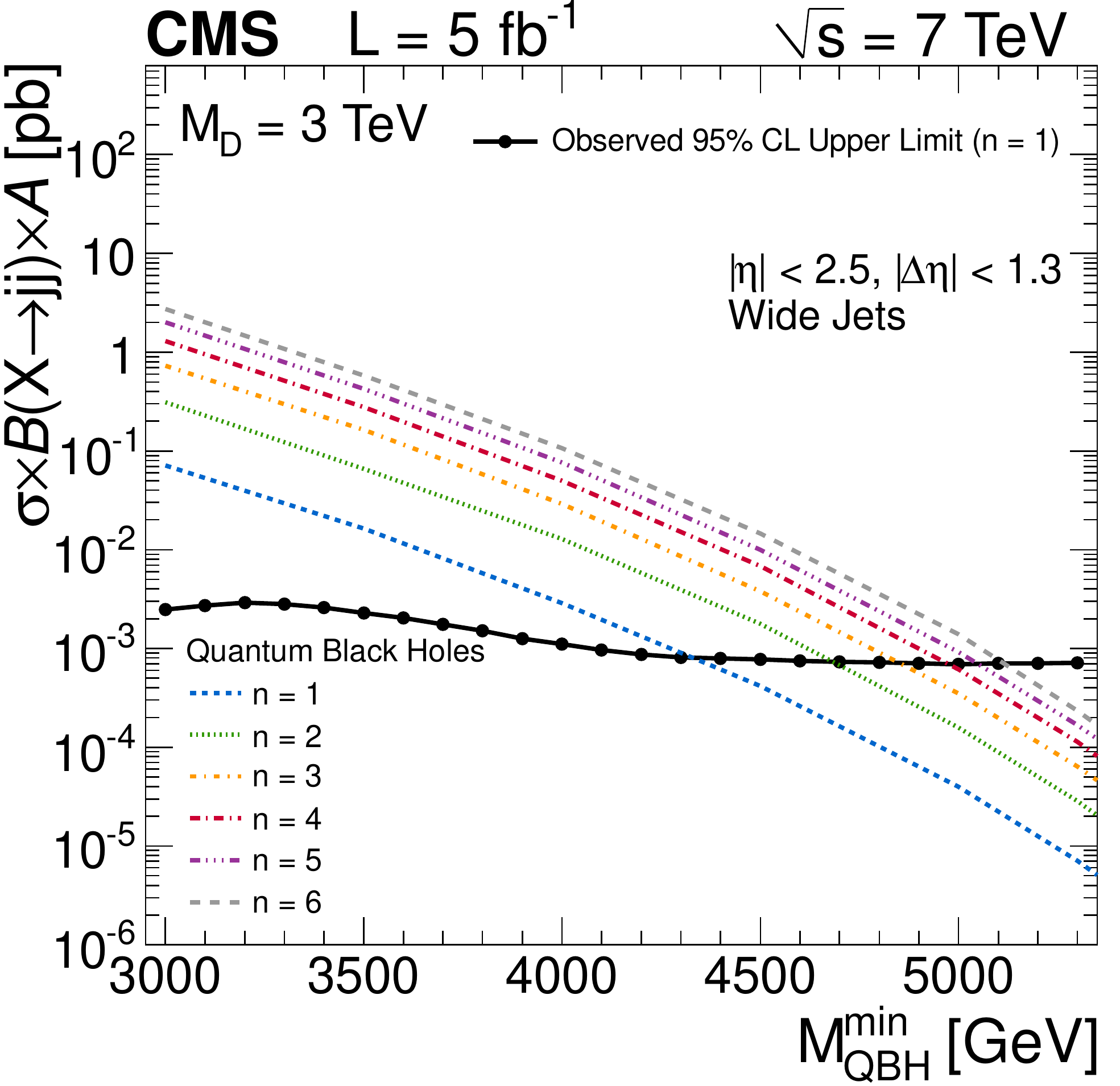} \\
   (a) & (b) \\
   \includegraphics[width=0.48\textwidth]{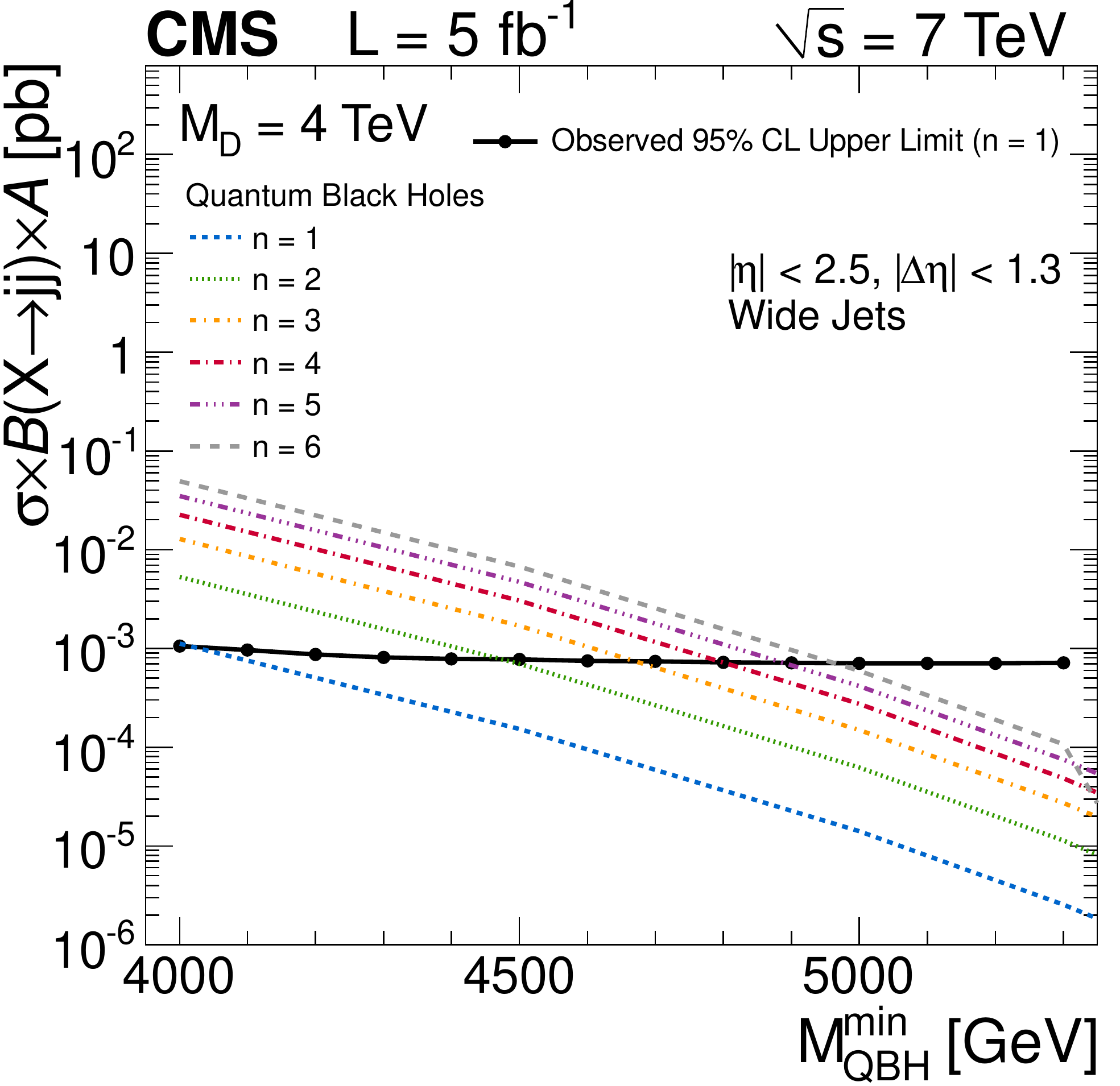} &
   \includegraphics[width=0.48\textwidth]{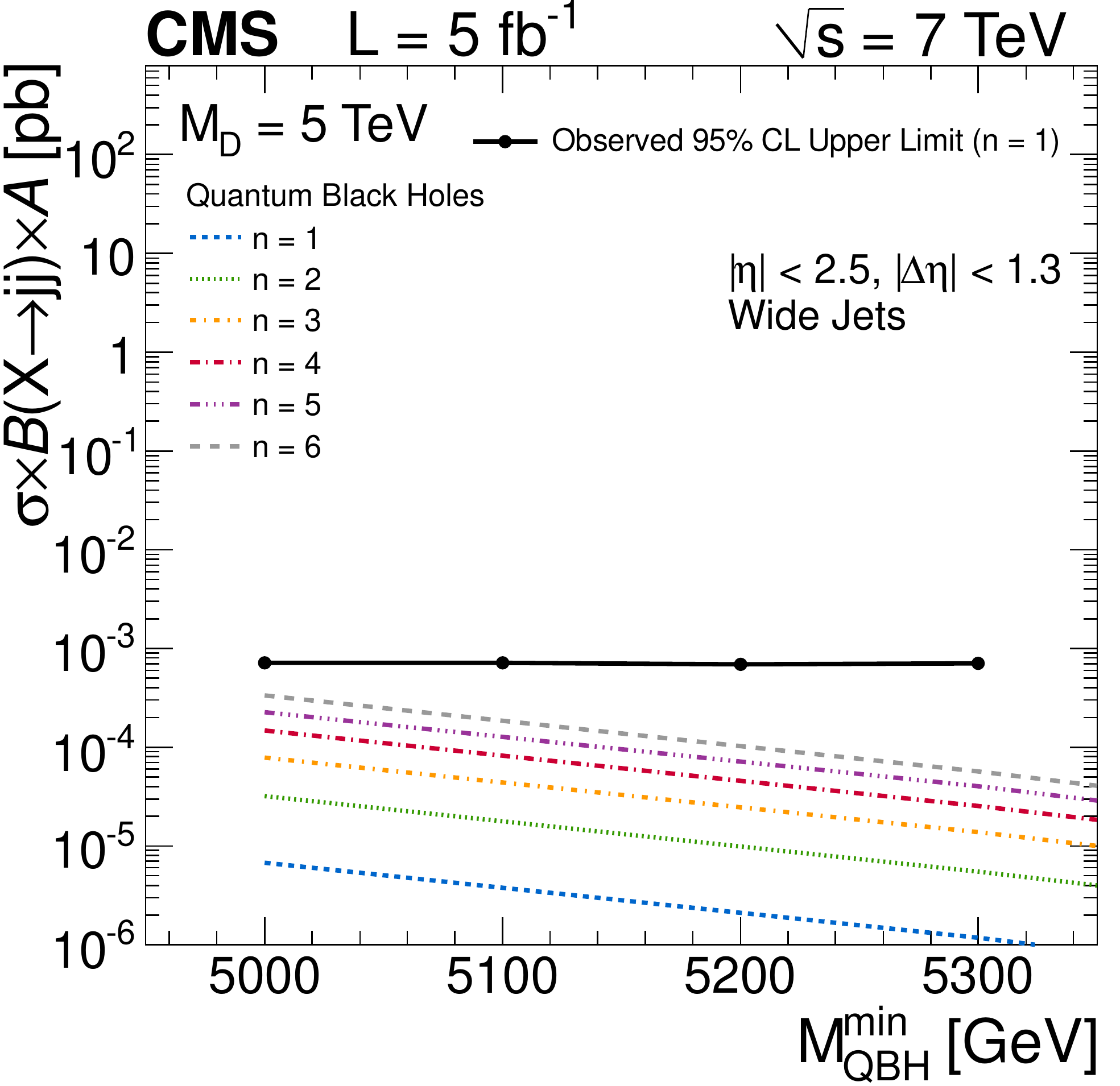} \\
   (c) & (d)
 \end{tabular}
  \caption{Observed 95\% CL upper limits on $\sigma\times B\times A$ as a function of the minimum mass of quantum
    black holes, compared to theoretical predictions for a quantum gravity scale of (a) $M_\text{D}=2$~TeV, (b)
    $M_\text{D}=3$~TeV, (c) $M_\text{D}=4$~TeV, and (d) $M_\text{D}=5$~TeV, with the number of extra dimensions $n$
    ranging from one to six. The observed upper cross section limits are fairly independent of $n$ (limits for $n=2$--$6$ are
    within ${\sim}5\%$ of those for $n=1$) and would be practically indistinguishable in the above plots; therefore, for
    display purposes, only the observed upper limits for $n=1$ are shown.
  \label{fig:qbh_limits}}
\end{figure}

\begin{figure}[!htb]
  \centering
  \includegraphics[width=0.48\textwidth]{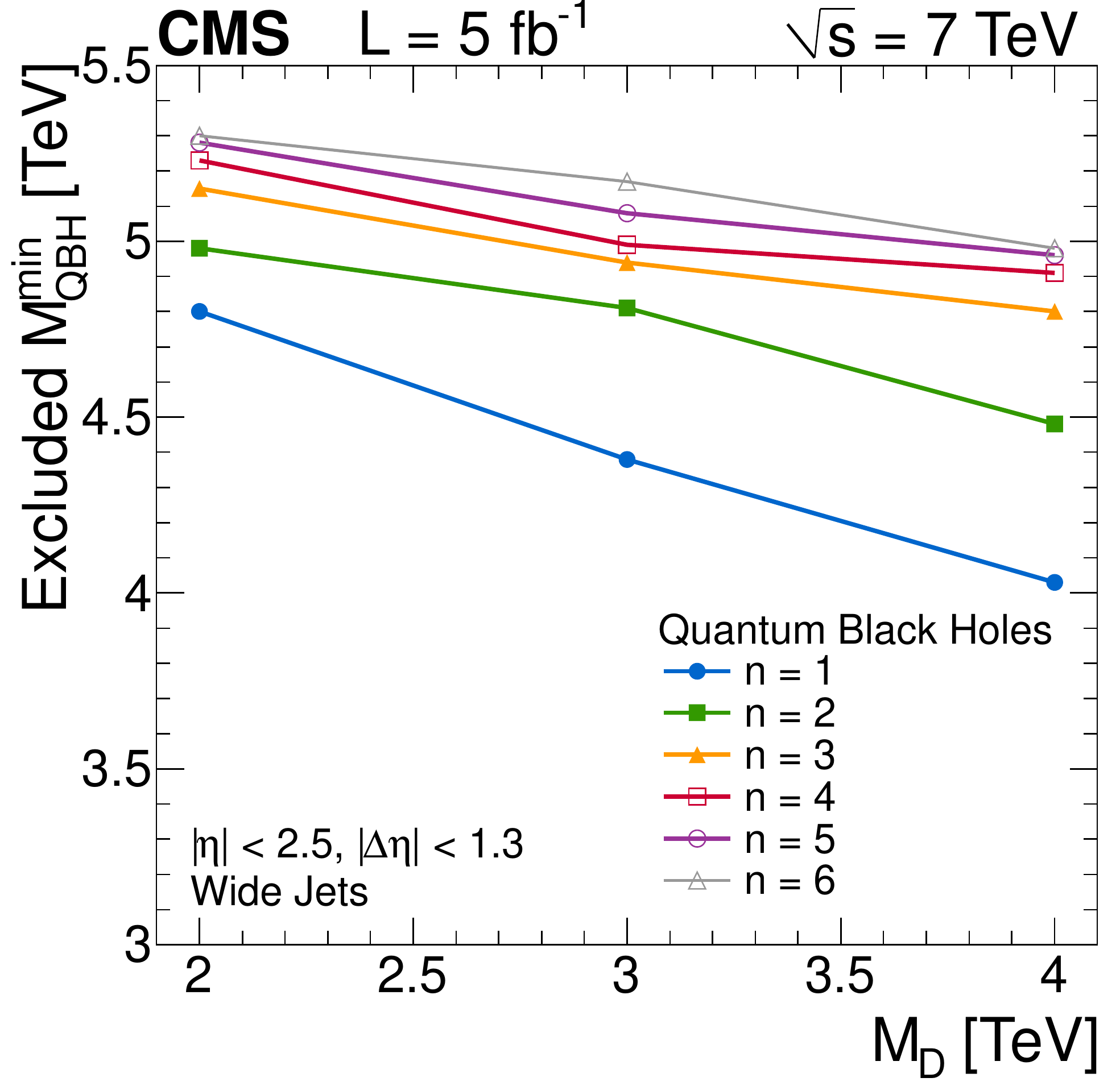}
  \caption{Observed 95\% CL lower limits on the minimum mass of quantum black holes as a function of the quantum gravity
   scale $M_\text{D}$ for the number of extra dimensions $n$ of one (Randall--Sundrum model) and two to six (ADD model).
  \label{fig:qbh_mass}}
\end{figure}

The 95\% CL observed upper limits on $\sigma\times B\times A$ for quantum black holes, derived from the inclusive analysis,
are shown in Fig.~\ref{fig:qbh_limits} and reported in Table~\ref{tab:limits_obs_inclusive_QBH}. The corresponding lower limits
on the minimum mass of quantum black holes range from 4 to 5.3~TeV, depending on the model parameters, and are shown in
Fig.~\ref{fig:qbh_mass} as a function of $M_\text{D}$. These limits are slightly better than those obtained in
Ref.~\cite{QBH2011}, where the same models were used. In Ref.~\cite{QBH2011}, a $S_\text{T}$ variable, defined as
$S_\text{T} = \sum\pt + \MET$ where the sum runs over individual objects: jets, electrons, photons, and muons, was used
as a discriminator between the signal and the background,
and counting experiments were performed above certain $S_\text{T}$ values. In this analysis we take advantage of the fact
that the shape of the signal in the dijet mass distribution is narrower than that in the generic $S_\text{T}$ variable.
This improved signal resolution allows us to extend the limits from the previous search. With the present data set, this
analysis is not yet sensitive to the production of quantum black holes with $M_\text{D}=5$~TeV and would require a factor
of $2$--$3$ increase in data to become sensitive to scenarios with $n=5$--$6$.

\begin{table}[!htbp]
 \caption{Observed 95\% CL upper limits on $\sigma\times B\times A$ for quantum black holes from the inclusive
  analysis. Only the limits for $n=1$ are reported. The limits for $n=2$--$6$ are within ${\sim}5\%$ of those for $n=1$.
 \label{tab:limits_obs_inclusive_QBH}}
 \centering
 \begin{tabular}{|c|c|c|c|c|} \hline
   $M_\text{QBH}^\text{min}$  &  \multicolumn{4}{c|}{Upper limit on $\sigma\times B\times A$ [pb]} \\
   $\,$[TeV]$\,$ &  $M_\text{D}=2$~TeV  &  $M_\text{D}=3$~TeV  &  $M_\text{D}=4$~TeV  &  $M_\text{D}=5$~TeV \\ \hline
   2.0 & 0.048   &         &         &         \\
   2.1 & 0.051   &         &         &         \\
   2.2 & 0.037   &         &         &         \\
   2.3 & 0.032   &         &         &         \\
   2.4 & 0.025   &         &         &         \\
   2.5 & 0.014   &         &         &         \\
   2.6 & 0.0061  &         &         &         \\
   2.7 & 0.0044  &         &         &         \\
   2.8 & 0.0035  &         &         &         \\
   2.9 & 0.0029  &         &         &         \\
   3.0 & 0.0026  & 0.0025  &         &         \\
   3.1 & 0.0027  & 0.0027  &         &         \\
   3.2 & 0.0030  & 0.0029  &         &         \\
   3.3 & 0.0028  & 0.0028  &         &         \\
   3.4 & 0.0025  & 0.0026  &         &         \\
   3.5 & 0.0022  & 0.0023  &         &         \\
   3.6 & 0.0020  & 0.0020  &         &         \\
   3.7 & 0.0017  & 0.0018  &         &         \\
   3.8 & 0.0015  & 0.0015  &         &         \\
   3.9 & 0.0013  & 0.0013  &         &         \\
   4.0 & 0.0010  & 0.0011  & 0.0011  &         \\
   4.1 & 0.00092 & 0.00096 & 0.00096 &         \\
   4.2 & 0.00087 & 0.00087 & 0.00087 &         \\
   4.3 & 0.00080 & 0.00081 & 0.00081 &         \\
   4.4 & 0.00077 & 0.00079 & 0.00078 &         \\
   4.5 & 0.00076 & 0.00077 & 0.00077 &         \\
   4.6 & 0.00074 & 0.00075 & 0.00075 &         \\
   4.7 & 0.00072 & 0.00073 & 0.00074 &         \\
   4.8 & 0.00072 & 0.00072 & 0.00072 &         \\
   4.9 & 0.00071 & 0.00071 & 0.00071 &         \\
   5.0 & 0.00071 & 0.00069 & 0.00071 & 0.00071 \\
   5.1 & 0.00069 & 0.00070 & 0.00071 & 0.00071 \\
   5.2 & 0.00070 & 0.00071 & 0.00070 & 0.00069 \\
   5.3 & 0.00068 & 0.00072 & 0.00071 & 0.00071 \\
 \hline
 \end{tabular}
\end{table}

\begin{figure}[!htb]
 \centering
 \begin{tabular}{cc}
   \includegraphics[width=0.48\textwidth]{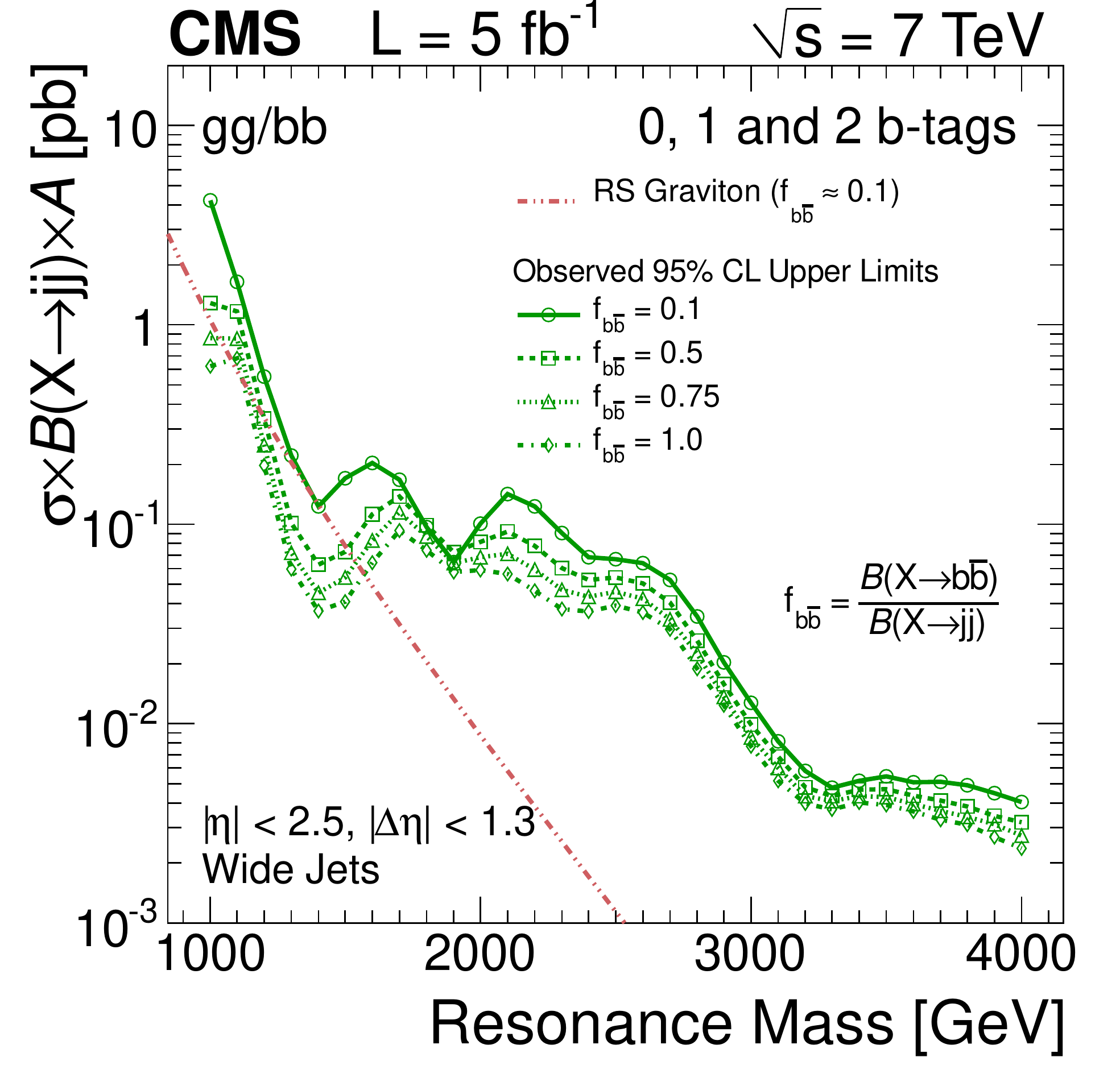} &
   \includegraphics[width=0.48\textwidth]{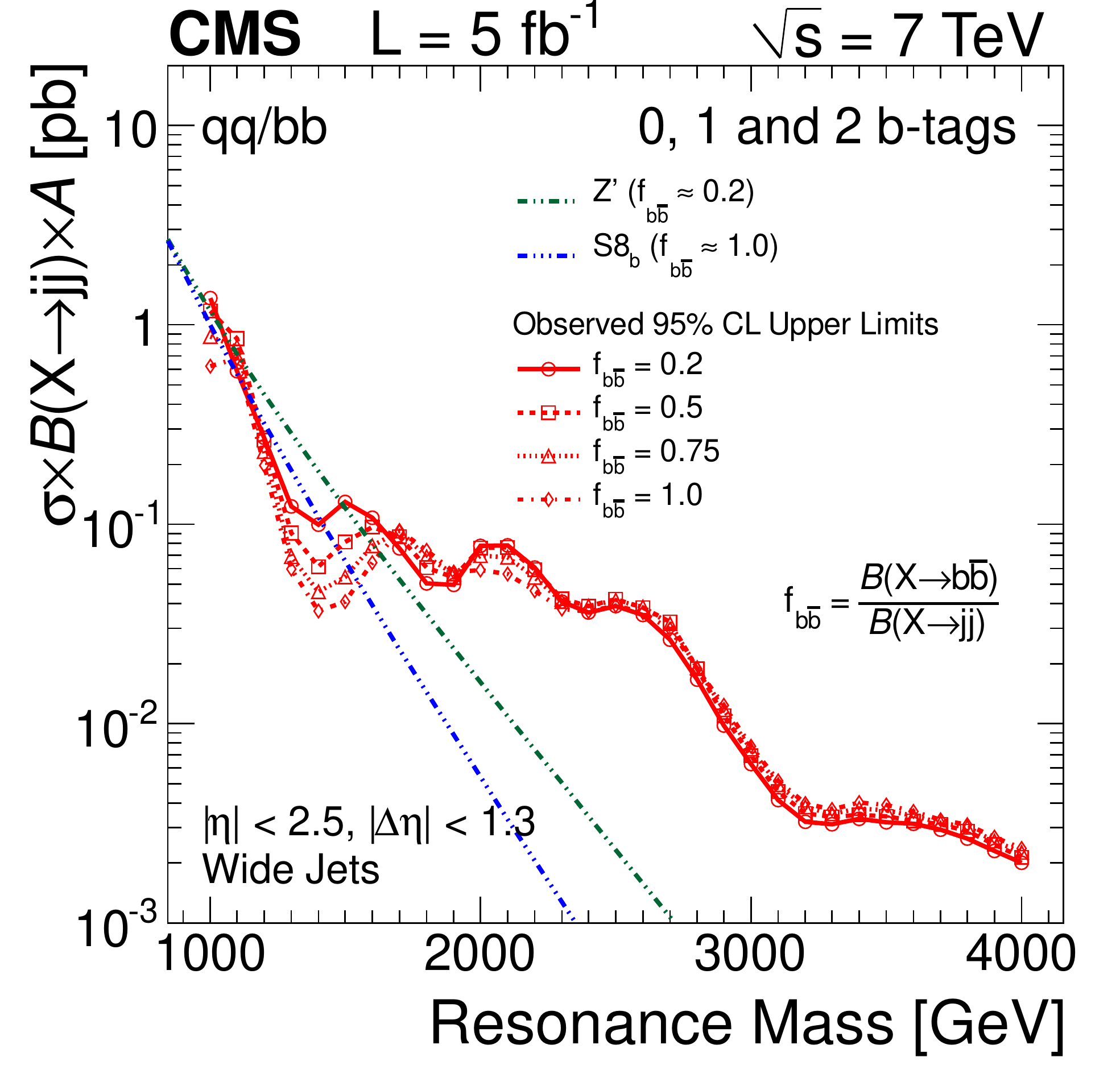} \\
   (a) & (b)
 \end{tabular}
 \caption{Observed 95\% CL upper limits on $\sigma\times B\times A$ for dijet resonances of type (a) gg/bb
    and (b) qq/bb, as defined in Section~\ref{sc:stat}, from the b-tagged analysis for four different values of $f_{\bbbar}$,
    compared to theoretical predictions for RS gravitons~\cite{ref_rsg}, $\cPZpr$ bosons~\cite{ref_gauge}, and S8$_\text{\cPqb}$
    resonances~\cite{Bai:2010dj}.
 \label{fig:limits_obs_btag_fbbbar}}
\end{figure}

Figure~\ref{fig:limits_obs_btag_fbbbar} shows the observed upper limits at the 95\% CL on $\sigma\times B\times
A$ for gg/bb and qq/bb resonances from the b-tagged analysis for different values of $f_{\bbbar}$. For any model with
known value of $f_{\bbbar}$, the prediction of $\sigma\times B\times A$ at the parton level has to be compared to
an appropriate limit curve in order to determine mass limits. The prediction for RS gravitons should be compared to the
$f_{\bbbar}=0.1$ limit curve, for $\cPZpr$ bosons to the $f_{\bbbar}=0.2$ limit curve, and for S8$_\text{\cPqb}$ resonances
to the $f_{\bbbar}=1.0$ limit curve. The observed upper limits for signal masses between 1.0 and 4.0~TeV and the values of
$f_{\bbbar}$ shown in Fig.~\ref{fig:limits_obs_btag_fbbbar} are reported in Table~\ref{tab:limits_obs_btag}. It is worth
noting that for gg/bb resonances, the limits become more stringent as $f_{\bbbar}$ increases. For example, for gg/bb
resonances with masses below $2$~TeV, the upper cross section limits are as much as $70\%$ lower for $f_{\bbbar}=1.0$ than for
$f_{\bbbar}=0.1$. For qq/bb resonances, however, this trend reverses at large values of the resonance mass since b
tagging starts to lose its discriminating power and qq/bb mass shapes become wider as $f_{\bbbar}$ increases.

\begin{table}[!htbp]
 \caption{Observed 95\% CL upper limits on $\sigma\times B\times A$ for narrow gg/bb and qq/bb resonances,
   as defined in Section~\ref{sc:stat}, from the b-tagged analysis for signal masses between 1.0 and 4.0~TeV.
 \label{tab:limits_obs_btag}}
 \centering
 \begin{tabular}{|c|c|c|c|c|c|c|c|} \hline
   Mass   &  \multicolumn{7}{c|}{Upper limit on $\sigma\times B\times A$ [pb]}\\
   $\,$[TeV]$\,$  &  \multicolumn{3}{c|}{gg/bb} & \multicolumn{3}{c|}{qq/bb}  &  gg/bb, qq/bb \\
          &  $f_{\bbbar}=0.1$  & $f_{\bbbar}=0.5$  &  $f_{\bbbar}=0.75$ & $f_{\bbbar}=0.2$ & $f_{\bbbar}=0.5$ & $f_{\bbbar}=0.75$ & $f_{\bbbar}=1.0$ \\ \hline
   1.0 & 4.2    & 1.3    & 0.86   & 1.4    & 1.2    & 0.87   & 0.62   \\
   1.1 & 1.6    & 1.2    & 0.85   & 0.59   & 0.85   & 0.78   & 0.68   \\
   1.2 & 0.55   & 0.34   & 0.25   & 0.27   & 0.26   & 0.23   & 0.20   \\
   1.3 & 0.22   & 0.10   & 0.072  & 0.12   & 0.09   & 0.069  & 0.060  \\
   1.4 & 0.12   & 0.063  & 0.045  & 0.099  & 0.061  & 0.046  & 0.037  \\
   1.5 & 0.17   & 0.073  & 0.054  & 0.13   & 0.082  & 0.055  & 0.041  \\
   1.6 & 0.20   & 0.11   & 0.083  & 0.11   & 0.097  & 0.078  & 0.064  \\
   1.7 & 0.17   & 0.14   & 0.11   & 0.076  & 0.087  & 0.093  & 0.093  \\
   1.8 & 0.096  & 0.099  & 0.087  & 0.051  & 0.061  & 0.069  & 0.074  \\
   1.9 & 0.065  & 0.072  & 0.064  & 0.050  & 0.054  & 0.056  & 0.058  \\
   2.0 & 0.10   & 0.082  & 0.068  & 0.078  & 0.076  & 0.069  & 0.059  \\
   2.1 & 0.14   & 0.092  & 0.071  & 0.078  & 0.077  & 0.068  & 0.056  \\
   2.2 & 0.12   & 0.078  & 0.059  & 0.061  & 0.059  & 0.054  & 0.046  \\
   2.3 & 0.09   & 0.060  & 0.047  & 0.041  & 0.042  & 0.041  & 0.038  \\
   2.4 & 0.068  & 0.052  & 0.043  & 0.036  & 0.039  & 0.039  & 0.037  \\
   2.5 & 0.067  & 0.054  & 0.046  & 0.039  & 0.042  & 0.042  & 0.039  \\
   2.6 & 0.064  & 0.050  & 0.042  & 0.035  & 0.038  & 0.038  & 0.036  \\
   2.7 & 0.053  & 0.040  & 0.033  & 0.026  & 0.032  & 0.032  & 0.030  \\
   2.8 & 0.035  & 0.026  & 0.022  & 0.017  & 0.019  & 0.019  & 0.019  \\
   2.9 & 0.020  & 0.016  & 0.014  & 0.0098 & 0.011  & 0.012  & 0.012  \\
   3.0 & 0.013  & 0.0099 & 0.0085 & 0.0063 & 0.0069 & 0.0075 & 0.0077 \\
   3.1 & 0.0082 & 0.0068 & 0.0060 & 0.0041 & 0.0046 & 0.0049 & 0.0052 \\
   3.2 & 0.0058 & 0.0048 & 0.0043 & 0.0032 & 0.0036 & 0.0038 & 0.0040 \\
   3.3 & 0.0048 & 0.0043 & 0.0041 & 0.0031 & 0.0034 & 0.0035 & 0.0037 \\
   3.4 & 0.0052 & 0.0047 & 0.0043 & 0.0033 & 0.0035 & 0.0037 & 0.0040 \\
   3.5 & 0.0054 & 0.0047 & 0.0043 & 0.0032 & 0.0034 & 0.0037 & 0.0039 \\
   3.6 & 0.0051 & 0.0043 & 0.0040 & 0.0031 & 0.0033 & 0.0034 & 0.0036 \\
   3.7 & 0.0051 & 0.0041 & 0.0037 & 0.0029 & 0.0031 & 0.0032 & 0.0033 \\
   3.8 & 0.0049 & 0.0038 & 0.0034 & 0.0026 & 0.0029 & 0.0030 & 0.0031 \\
   3.9 & 0.0045 & 0.0034 & 0.0031 & 0.0023 & 0.0025 & 0.0026 & 0.0027 \\
   4.0 & 0.0041 & 0.0032 & 0.0027 & 0.0020 & 0.0021 & 0.0023 & 0.0024 \\
 \hline
 \end{tabular}
\end{table}

\begin{figure}[!htb]
 \centering
 \begin{tabular}{cc}
   \includegraphics[width=0.48\textwidth]{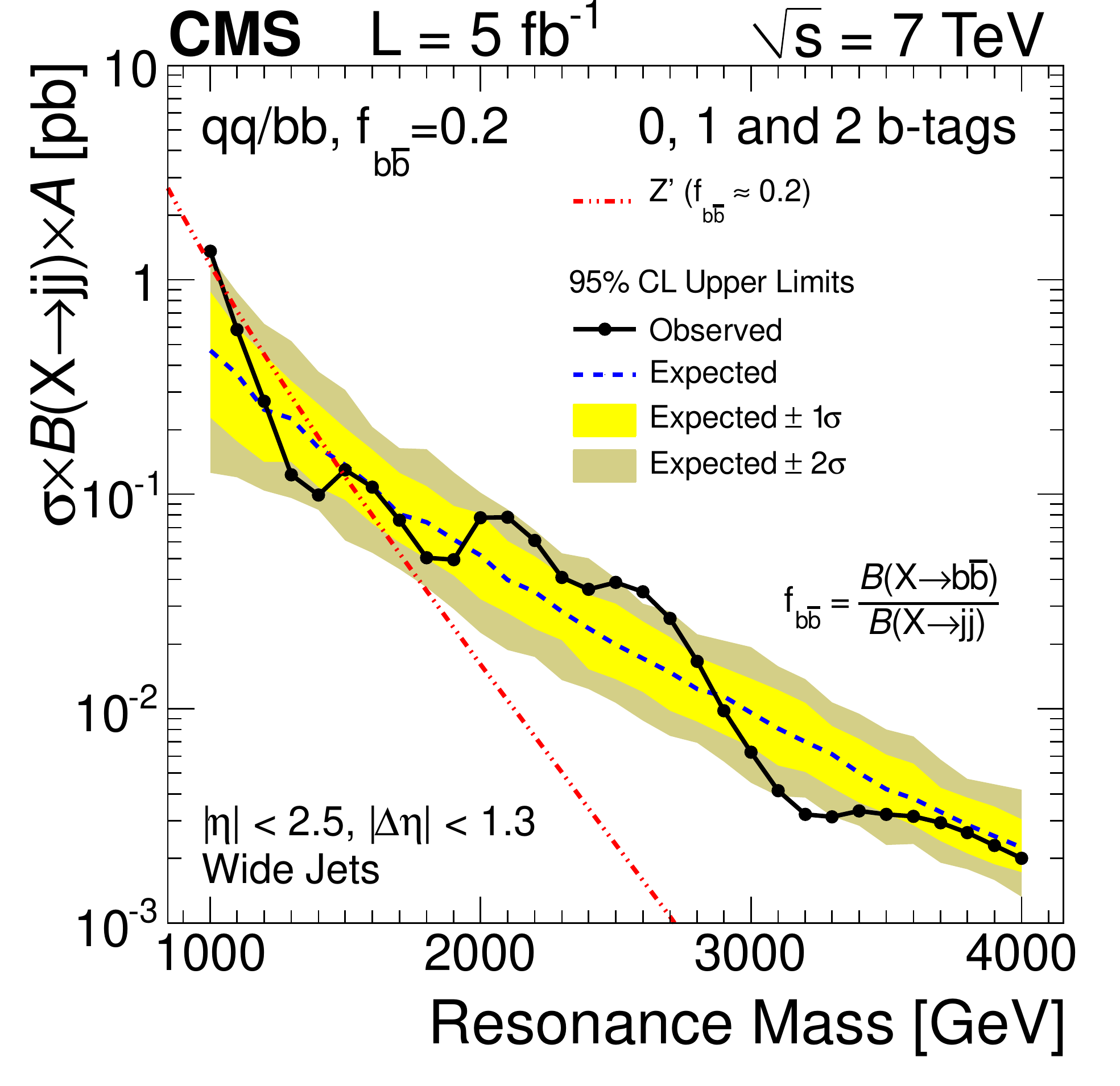} &
   \includegraphics[width=0.48\textwidth]{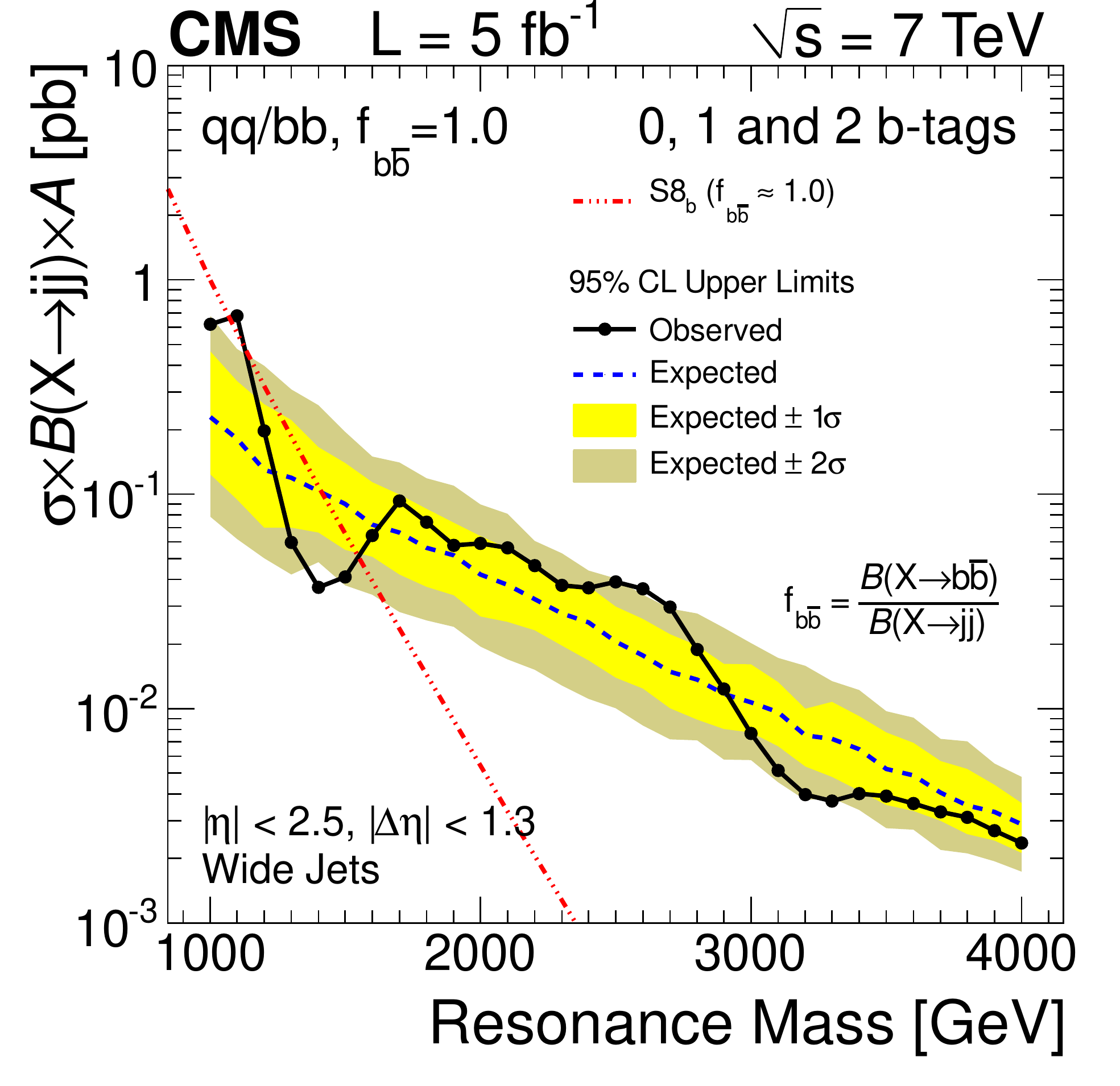} \\
   (a) & (b)
 \end{tabular}
 \caption{Observed 95\% CL upper limits on $\sigma\times B\times A$ for qq/bb resonances, as defined in
   Section~\ref{sc:stat}, from the b-tagged analysis with (a) $f_{\bbbar}=0.2$ and (b) $f_{\bbbar}=1.0$ (points),
   compared to the expected limits (dashed) and their variation at 1$\sigma$ and 2$\sigma$ levels (shaded bands).
   Theoretical predictions for $\cPZpr$ bosons and S8$_\text{\cPqb}$ resonances are also shown.
 \label{fig:limits_exp_btag_Zprime_S8b}}
\end{figure}

Figure~\ref{fig:limits_exp_btag_Zprime_S8b} shows the expected limits and their uncertainty bands for qq/bb resonances
with $f_{\bbbar}=0.2$ and $f_{\bbbar}=1.0$ and compares them to both the observed limits and theoretical predictions for
$\cPZpr$ bosons and S8$_\text{\cPqb}$ resonances. The expected exclusion for $\cPZpr$ bosons is $1.0 < M(\cPZpr) < 1.45$~TeV,
and we exclude $1.04 < M(\cPZpr) < 1.49$~TeV. For S8$_\text{\cPqb}$ resonances, the expected exclusion is
$1.0 < M(\text{S8}_\text{b}) < 1.42$~TeV, and we exclude $1.0 < M(\text{S8}_\text{b}) < 1.08$~TeV and
$1.12 < M(\text{S8}_\text{b}) < 1.56$~TeV. With the present data, no limits are set on the RS graviton mass. The
observed and expected mass exclusions from the b-tagged analysis for $\cPZpr$ bosons and S8$_\text{\cPqb}$ resonances are
summarized in Table~\ref{tab:mass_exclusions_btag}.

\begin{table}[!htb]
  \caption{Observed and expected 95\% CL mass exclusions from the b-tagged analysis for $\cPZpr$ bosons and
    S8$_\text{\cPqb}$ resonances.}
  \centering
  \normalsize
  \begin{tabular}{|c|c|c|c|}
  \hline
    Model & Final State & Exp. Mass Exclusion & Obs. Mass Exclusion \\
          &             & [TeV]     & [TeV] \\
    \hline
    $\cPZpr$ Boson ($\cPZpr$)                       & $\cPq\cPaq$ & [1.0, 1.45] & [1.04, 1.49] \\
    S8$_\text{\cPqb}$ Resonance (S8$_\text{\cPqb}$) & $\cPq\cPaq$ & [1.0, 1.42] & [1.0, 1.08], [1.12, 1.56] \\
    \hline
  \end{tabular}
\label{tab:mass_exclusions_btag}
\end{table}

\section{Summary}

A search for narrow resonances and quantum black holes in the dijet mass spectra has been performed
using pp collisions at $\sqrt{s}=7$~TeV, collected by the CMS detector at the LHC. Measured dijet mass spectra with and
without b-tagging requirements are observed to be consistent with the standard model expectation of a smoothly falling distribution.
There is no evidence for new particle production in the data. Model-independent upper limits are presented
on the product $\sigma\times B\times A$ that are applicable to any model of narrow dijet resonance production,
and with b tagging applied, limits are expressed in terms of the branching fraction to b-jet pairs. Lower limits are obtained
on the mass of string resonances, excited quarks, axigluons and colorons, scalar color-octet resonances,
E$_{6}$ diquarks, W$^\prime$ and Z$^\prime$ bosons, and quantum black holes. Most of these limits extend excluded mass ranges
from the previous searches.

\section*{Acknowledgements}

We thank Can Kilic for calculations of the string resonance cross section, Ian Lewis for calculations of
the S8 model cross section, and Bogdan Dobrescu for assistance in implementing the S8$_\text{\cPqb}$ model.
We congratulate our colleagues in the CERN accelerator departments for the excellent performance of the LHC machine. We
thank the technical and administrative staff at CERN and other CMS institutes, and acknowledge support from: BMWF and
FWF (Austria); FNRS and FWO (Belgium); CNPq, CAPES, FAPERJ, and FAPESP (Brazil); MES (Bulgaria); CERN; CAS, MoST, and
NSFC (China); COLCIENCIAS (Colombia); MSES (Croatia); RPF (Cyprus); MoER, SF0690030s09 and ERDF (Estonia); Academy of
Finland, MEC, and HIP (Finland); CEA and CNRS/IN2P3 (France); BMBF, DFG, and HGF (Germany); GSRT (Greece); OTKA and NKTH
(Hungary); DAE and DST (India); IPM (Iran); SFI (Ireland); INFN (Italy); NRF and WCU (Korea); LAS (Lithuania);
CINVESTAV, CONACYT, SEP, and UASLP-FAI (Mexico); MSI (New Zealand); PAEC (Pakistan); MSHE and NSC (Poland); FCT
(Portugal); JINR (Armenia, Belarus, Georgia, Ukraine, Uzbekistan); MON, RosAtom, RAS and RFBr (Russia); MSTD (Serbia);
SEIDI and CPAN (Spain); Swiss Funding Agencies (Switzerland); NSC (Taipei); TUBITAK and TAEK (Turkey); STFC (United
Kingdom); DOE and NSF (USA).

\bibliography{auto_generated}   % will be created by the tdr script.

\cleardoublepage \appendix\section{The CMS Collaboration \label{app:collab}}\begin{sloppypar}\hyphenpenalty=5000\widowpenalty=500\clubpenalty=5000\textbf{Yerevan Physics Institute,  Yerevan,  Armenia}\\*[0pt]
S.~Chatrchyan, V.~Khachatryan, A.M.~Sirunyan, A.~Tumasyan
\vskip\cmsinstskip
\textbf{Institut f\"{u}r Hochenergiephysik der OeAW,  Wien,  Austria}\\*[0pt]
W.~Adam, E.~Aguilo, T.~Bergauer, M.~Dragicevic, J.~Er\"{o}, C.~Fabjan\cmsAuthorMark{1}, M.~Friedl, R.~Fr\"{u}hwirth\cmsAuthorMark{1}, V.M.~Ghete, N.~H\"{o}rmann, J.~Hrubec, M.~Jeitler\cmsAuthorMark{1}, W.~Kiesenhofer, V.~Kn\"{u}nz, M.~Krammer\cmsAuthorMark{1}, I.~Kr\"{a}tschmer, D.~Liko, I.~Mikulec, M.~Pernicka$^{\textrm{\dag}}$, D.~Rabady\cmsAuthorMark{2}, B.~Rahbaran, C.~Rohringer, H.~Rohringer, R.~Sch\"{o}fbeck, J.~Strauss, A.~Taurok, W.~Waltenberger, C.-E.~Wulz\cmsAuthorMark{1}
% ...
\vskip\cmsinstskip
\dag:~Deceased\\
1:~~Also at Vienna University of Technology, Vienna, Austria\\
2:~~Also at CERN, European Organization for Nuclear Research, Geneva, Switzerland\\
% ...

\end{sloppypar}
\end{document}